\documentclass[11pt]{article}

\usepackage{amssymb}
\usepackage{amsthm}
\usepackage{amsmath}

\usepackage{subfigure}
\usepackage{graphicx}
\usepackage{psfrag}

\def\be{\begin{equation}}
\def\ee{\end{equation}}
\def\bea{\begin{eqnarray}}
\def\eea{\end{eqnarray}}
\def\lb{\label}

\newcommand{\sfrac}[2]{{\textstyle{\frac{#1}{#2}}}}
\newcommand{\textfrac}[2]{{\textstyle{\frac{#1}{#2}}}}
\newcommand{\ce}{{\mathcal E}}
\newcommand{\cl}{{\mathcal L}}

\begin{document}


\title{\sc Matter and dynamics in closed cosmologies}
\author{\sc
J.\ Mark Heinzle$^{1}$\thanks{Electronic address: {\tt
Mark.Heinzle@aei.mpg.de}}\ ,\ Niklas
R\"ohr$^{2}$\thanks{Electronic address: {\tt Niklas.Rohr@kau.se}}\
,\ \ and Claes Uggla$^{2}$\thanks{Electronic address:
{\tt Claes.Uggla@kau.se}}\\
$^{1}${\small\em Max-Planck-Institute for Gravitational Physics, Albert-Einstein-Institute,}\\
{\small\em D-14 476 Golm, Germany}\\
$^{2}${\small\em Department of Physics, University of Karlstad,}\\
{\small\em S-651 88 Karlstad, Sweden}}

\date{\normalsize{June 18, 2004}}
\maketitle

\begin{abstract}

To systematically analyze the dynamical implications of the matter
content in cosmology, we generalize earlier dynamical systems
approaches so that perfect fluids with a general barotropic
equation of state can be treated. We focus on locally rotationally
symmetric Bianchi type IX and Kantowski-Sachs orthogonal perfect
fluid models, since such models exhibit a particularly rich
dynamical structure and also illustrate typical features of more
general cases. For these models, we recast Einstein's field
equations into a regular system on a compact state space, which is
the basis for our analysis. We prove that models expand from a
singularity and recollapse to a singularity when the perfect fluid
satisfies the strong energy condition. When the matter source
admits Einstein's static model, we present a comprehensive
dynamical description, which includes asymptotic behavior, of
models in the neighborhood of the Einstein model; these results
make earlier claims about ``homoclinic phenomena and chaos''
highly questionable. We also discuss aspects of the global
asymptotic dynamics, in particular, we give criteria for the
collapse to a singularity, and we describe when models expand
forever to a state of infinite dilution; possible initial and
final states are analyzed. Numerical investigations complement the
analytical results.

\end{abstract}
\centerline{\bigskip\noindent PACS number(s): 04.20-q, 98.80.Jk,
98.80.Cq, 04.20.Ha} \vfill
\newpage

\section{Introduction}
\label{introduction}

At present, the matter content in the Universe is a mystery: dark
matter, dark energy, quintessence; our knowledge that underlies
these labels is clearly not as good as we would wish. Apart from
observational challenges this also poses theoretical issues, for
example, very little is known about how matter properties
influence cosmological evolution, the character of singularities,
and the stability of various special cosmological models. This
motivates the development of a framework that is insensitive to
the details of the matter content, but still allows one to
systematically investigate the relationship between properties of
matter and cosmological dynamics. The book edited by Wainwright
and Ellis~\cite{book:waiell97} used dynamical systems methods to
successfully study perfect fluid models with a linear equation of
state; in this paper we modify and generalize the formalism
in~\cite{book:waiell97}, and attempt to show that the usefulness
of dynamical systems methods increases with the complexity of the
matter source.

In this paper we consider spatially homogeneous (SH) cosmological
models with orthogonal perfect fluid matter (i.e., the fluid
velocity is orthogonal with respect to the symmetry surfaces)
described by an effective barotropic equation of state $p =
p(\rho)$, where $p$ denotes the pressure and $\rho$ the
energy-density, which we assume to be non-negative; the effective
fluid may consist of an arbitrary number of non-interacting
orthogonal perfect fluids with non-negative energy-densities and
barotropic equations of state, and a positive cosmological
constant. From these assumptions it follows that SH Bianchi type
I-VIII cosmologies are forever expanding or contracting and
comparatively easy to treat, as discussed in the concluding
remarks. We therefore focus on SH Bianchi type IX and
Kantowski-Sachs perfect fluid models which exhibit a particularly
rich dynamical structure. These models are the only SH models that
admit positive intrinsic curvature; they thus naturally have a
closed spatial topology and we therefore refer to them as closed
cosmologies (although all Bianchi class A and some class B models
also admit closed spatial topologies). However, we restrict
ourselves to the locally rotationally symmetric (LRS) orthogonal
perfect fluid cases, since this suffices to illustrate our
approach; generalizations, even to models with no symmetries at
all, are discussed in the concluding remarks.

The LRS type IX and Kantowski-Sachs models have been studied
before: in~\cite{uggmuhl990}, \cite{golell1999}, \cite{colgol2000}
dynamical systems techniques were used to study quite special
sources, e.g., perfect fluids with a linear equations of state and
a cosmological constant; other methods were used
in~\cite{Lin/Wald:1990} and~\cite{Rendall:1995} to treat more
general matter that included perfect fluids with non-negative
pressure, note however that this excludes e.g., the inclusion of a
cosmological constant. Recently the dynamics associated with
Einstein's static model was studied in a series of papers that
covered several special cases in a LRS type IX context: a positive
cosmological constant and dust in \cite{Oliveira/etal:1997}, a
positive cosmological constant and a perfect fluid with a linear
equation of state in \cite{Barguine/etal:2001}; the latter work
was subsequently generalized to the diagonal non-LRS type IX case
in \cite{olietal2002}. In these papers it was asserted that the
presence of Einstein's static model leads to the existence of
infinitely many homoclinic orbits [orbits whose $\alpha$- and
$\omega$-limit is the same periodic orbit] that ``produce chaotic
sets in the whole state space''; the asymptotic configurations of
models in a neighborhood of Einstein's model, as of any domain in
the whole phase space, were claimed to be unpredictable and
characterized by ``homoclinic chaos.'' We reach a different
conclusion: we present a comprehensive description of the dynamics
in the neighborhood of the Einstein model, moreover, we establish
theorems about the global asymptotic behavior of solutions. Our
results disprove the claims of non-predictability and chaos for
models close to the Einstein model; furthermore, we prove that the
asserted ``homoclinic phenomena'' must be confined to quite small
regions of the phase space, if they occur at all; indeed, our
numerical investigations suggest that ``homoclinic chaos'' is
excluded entirely, but further studies are needed to establish
this with certainty.

Our investigation is based on a reformulation of Einstein's field
equations for LRS type IX
(Kantowski-Sachs) models as a completely regular dynamical
system on a compact four-dimensional (three-dimensional) state
space. Building on this,
our analysis of the type IX case rests
on three cornerstones: methods from dynamical systems theory, in
particular center manifold analysis, the existence of a conserved
``energy'' integral, and the close relationship between the
dynamics of the full system and the Friedmann-Robertson-Walker
(FRW) case.

The outline of the paper is as follows: in Sec.~\ref{dynsysappro},
we reformulate Einstein's field equations for the LRS type IX and
Kantowski-Sachs models as a regular dynamical system on a compact
state space; in type IX we also find a conserved ``energy.'' In
Sec.~\ref{FRWSec} we describe the FRW models; first with a
potential approach, presumably familiar to a general reader, and
then with our dynamical systems formulation. In contrast to the
potential formulation, the dynamical systems approach can be
applied to more general cases. In Sec.~\ref{LRSIX} we investigate
the LRS type IX models. Together with dynamical systems methods
and center manifold reduction we use the conserved quantity to
obtain comprehensive information about the dynamical
possibilities. The dynamical system contains the Kantowski-Sachs
models as an invariant boundary subset; these models are briefly
treated in Sec.~\ref{KSmodels}. We conclude with a discussion
about other cases and how the present work fits in a larger
context as regards the dynamics of quite general models in
Sec.~\ref{discandout}. Some proofs and supplementary theorems
about aspects of the global dynamics are postponed to the
appendix, where we also show how our variables are related to the
Hubble-normalized variables used in e.g.,~\cite{book:waiell97}.
Throughout we use units such that the speed of light $c$ and
Newton's gravitational constant $G$ are given by $c=1=8\pi G$.

\section{The dynamical systems approach}
\label{dynsysappro}

The LRS Bianchi type IX models are characterized by the equations
\begin{alignat}{3} \dot{H} & =
-H^2-2\sigma_+^2-\sfrac{1}{6}(\rho+3p)\, ,& \qquad  \dot{\sigma}_+
&=
-3H\sigma_+ - {^3}S_+\, , \nonumber\\
\lb{dim*}\dot{n}_1 & =
-(H+4\sigma_+)n_1\, ,& \qquad \dot{n}_2  & = -(H-2\sigma_+)n_2\, , \\
 \dot{\rho} & = -3(\rho + p)H\, ,& \qquad \rho & =
3H^2-3\sigma_+^2+\sfrac{1}{2}{^3}R\ ,\nonumber
\end{alignat}
(see p.~40, 123--124 in \cite{book:waiell97}; $n_2=n_3$,
$\sigma_-=0$ yields the LRS case). The Hubble scalar $H$ allows
one to define the length scale $\ell$ through $H=\dot{\ell}/\ell$,
where $\ell$ is the cubic root of the volume density. The shear is
determined by $\sigma_+$; $n_1$ and $n_2$ describe the spatial
three-curvature via ${}^3R = -\textfrac{1}{2}n_1^2 + 2n_1n_2$ and
$S_+ = \frac{1}{3}(n_1n_2-n_1^2)$. The overdot denotes
differentiation w.r.t.~clock time $t$.

The algebraic equation for $\rho$ yields $H^2 + \frac{1}{3}n_1n_2
= \frac{1}{3}\rho + \sigma_+^2 + \frac{1}{12}n_1^2$. Since $n_1$
and $n_2$ have the same sign in Bianchi type IX, we can replace
$n_2$ with the variable $D:= (H^2+ n_1 n_2/3)^{1/2}$\,.
We then introduce the bounded dimensionless%
    \footnote{For a
      discussion about dimensions see, e.g., \cite{uggetal2004}.}
variables
\begin{alignat}{2}\lb{defvar}
\left(Q_0,\ Q_+ \right) & := \left(H,\ \sigma_+ \right)/D\, ,&
\qquad (M_1,\Omega_D) := \left(n_1^2,4\rho\right)/(12D^2)\, .
\end{alignat}
By definition, $-1<Q_0<1$ and $M_1>0$, while $\rho >0$ implies
\begin{equation}
\lb{OmDcon} \Omega_D= 1-Q_+^2-M_1 > 0\, .
\end{equation}
We replace $H,\sigma_+, n_1, n_2$ with $Q_0,Q_+,M_1,D$, and $\rho$
with $\Omega_D$, which is given in terms of $Q_+$ and $M_1$
by~(\ref{OmDcon}); instead of $p$ we introduce the dimensionless
quantity $w=p/\rho$. Then~(\ref{dim*}) yields one decoupled
dimensional equation
\be \lb{decoup} D' = -D(Q_0 F + Q_0 + Q_+)\, ,\qquad F := 2Q_+^2 -
Q_0Q_+ + \sfrac{1}{2}(1+3w(L))\Omega_D\, , \ee
and the coupled dimensionless system
\begin{equation}\label{dynsys}
\begin{array}{lll}
Q_0^\prime & = &-(1-Q_0^2)F\, , \\[0.7ex]
Q_+^\prime & = & 4M_1 - 1 +
(Q_0 - Q_+)^2 + Q_0 Q_+ F\, , \\[0.7ex]
M_1^\prime & = & 2M_1(Q_0 F - 3Q_+ )\, , \\[0.7ex]
L^\prime & = & f(L)L(1-L)Q_0\, ,
\end{array}
\end{equation}
where ${}^\prime =\frac{1}{d\tau}:=\frac{1}{Ddt}$ stems from a new
dimensionless time variable $\tau$; the variable $L$ is discussed
below.
The system~(\ref{dynsys}) is complemented by the auxiliary
equation
\begin{equation}\label{Omwprime}
\Omega_{D}^\prime =   (2Q_0 F - (1+3 w)Q_0 + 2Q_+)\Omega_{D}\:.
\end{equation}

The coupled system~(\ref{dynsys}) describes the essential
dynamics; once solved, the metric can be obtained in terms of
quadratures, cf.~Ch.~10 in~\cite{book:waiell97}. The r.h.s.~is
polynomial in $Q_0,Q_+,M_1$. This makes an inclusion of the
boundary, which consists of invariant subsets, possible; $Q_0=\pm
1$ yields the forever expanding and contracting representations of
the LRS Bianchi type II models; $M_1=0$ yields the Kantowski-Sachs
models (see, e.g.,~\cite{ashetal1993}) and $\Omega_D= 0$ the
vacuum boundary.
We denote the state space of~(\ref{dynsys}) by
\begin{equation}\label{4X}
{}^4\textbf{X}=\{(Q_0,Q_+,M_1,L)\}\, .
\end{equation}

The choice of the variable $L$ depends on the equation of state;
$L$ is constructed from $\ell$ in two steps: (i) Let
$x:=\ell/\ell_0$, where $\ell_0=\ell(t_0)$ and $t_{0}$ is a fixed
reference time; introduce $\lambda(x)$ subject to the conditions
$\lambda(x) \in \mathcal{C}^0[0,\infty)$, $\lambda(x) \in
\mathcal{C}^2(0,\infty)$, $d\lambda/dx>0$, $\lambda(0)=0$, and
$\lim_{x\rightarrow\infty}\lambda(x)\rightarrow\infty$. (ii)
Introduce $L:= \lambda/(1+\lambda)$; it follows that $L(x)$ is
monotonically increasing, and $L\in (0,1)$ for $x>0$, thus
$L\rightarrow 0$ describes a singularity while $L\rightarrow 1$
represents a state of infinite dispersion. Moreover, we define
$f(L):= {d\ln\lambda}/{d \ln x}|_{x(L)}$.

The function $w$ in~(\ref{decoup}) is to be regarded as a function $w(L)$:
from~(\ref{dim*}) we have $d\rho/(\rho+p) = -3 d x/x$, which
can be integrated to yield a function $\rho(x)$ when an equation
of state $p(\rho)$ is given, see also the example below.
Via $x= x(L)$ this leads to $w(L):=p/\rho\,|_{x(L)}$.

The variables $n_{1,2}$ are algebraically related to the spatial
metric and $H$, $\sigma_+$ to the extrinsic curvature (see Ch.~10
in~\cite{book:waiell97}), and are hence continuous in $t$, or
equivalently in $x$ and $L$; $\rho$ is continuous by virtue
of~(\ref{dim*}), $Q_0,Q_+,M_1,D$ are continuous
through~(\ref{defvar}); however, $p$, and hence $w = p/\rho$,
is not necessarily continuous -- jumps in $p$ correspond to phase
transitions. Although these can be covered by our formalism, we
will for simplicity assume that $p(\rho)$ is at least ${\mathcal
C}^1$; therefore $w(L)$ and thus the r.h.s.~of~(\ref{dynsys}) is
${\mathcal C}^1(0,1)$ (see~\cite{heiugg2003} for related issues).

For large classes of equations of state, $L$ can be chosen so
that~(\ref{dynsys}) is endowed with the desirable
differentiability properties for all $L\in [0,1]$, and thus the
boundaries $L=0,1$ can be included. We define an equation of state
to be {\em asymptotically linear\/} if $w\rightarrow dp/d\rho
\rightarrow const\, (=w_0, w_1)$ when $p\rightarrow 0,\infty$. In
this case we choose $L$ so 
that $w(L)$ and $f(L)$ become $\mathcal{C}^1$ on $L \in [0,1]$;
thus $w(0)=w_0$, $w(1)=w_1$, $f(0)=f_0$, $f(1)=f_1$. We restrict
ourselves to the situation where $f_0 > 0$ and $f_1 > 0$ is
possible.%
    \footnote{By these conditions excessive use of center manifold
      theory in the dynamical systems analysis is avoided.
      See~\cite{heietal2003} for a related problem where the
      general situation is discussed.}

To make things less abstract, let us consider
the special case of a source that consists
of an arbitrary number
of non-interacting perfect fluids with linear equations of state;
the total energy-density is
$\rho(x)=\sum_i\rho_{i0}x^{-3(1+w_i)}$, where $\sum_i\rho_{i0} =
\rho_0$ (the total energy-momentum $T^{\mu\nu}$ is a sum of the
$T_i^{\mu\nu}$ for each fluid component $i$; each component
satisfies ${T^{\mu\nu}_{i}}{}_{;\nu}=0$, and hence
$T^{\mu\nu}{}_{;\nu}=0$). We assume that $\rho_{i0} \geq 0$ and
$w_i \geq -1\:\,\forall\, i$; causality of each component requires
$w_i\leq 1$. As an example of how to obtain a suitable variable
$L$, let us take the combination of two fluids with $w_0>w_1$; the
choice $\lambda(x)=\frac{\rho_{10}}{\rho_{00}} x^{3(w_0-w_1)}$
yields $w = w_0 -(w_0-w_1)L$ and $f(L)=3(w_0-w_1)$, which clearly
satisfies the above conditions. As a second example, consider a
source that consists of radiation, dust and a cosmological
constant (a cosmological constant corresponds to
${\rho}_{\Lambda0}=\Lambda$ and $w_\Lambda=-1$); a possible choice
that satisfies the requirements is $\lambda=x$, which implies
$f(L)=1$ and
\begin{equation}\label{LdrEOS}
w(L)=\frac{\textfrac{1}{3}{\rho}_{r 0}\,(1-L)^4 - {\rho}_{\Lambda 0}
L^4}{{\rho}_{r 0}\,(1-L)^4 + {\rho}_{d 0}\,L\,(1-L)^3 +
{\rho}_{\Lambda 0}\,L^4}\, .
\end{equation}

The dimensionless coupled dynamical system~(\ref{dynsys}) is thus
regular everywhere on the state space ${}^4\textbf{X}$, including
the boundaries $Q_0= \pm 1$, $Q_+ =\pm 1$, $M_1=0$, and $\Omega_D
=0$; $L=0$, $L=1$ are also included when the equation of state is
asymptotically linear, however, even when this is not the case,
attractors and repellors are distinct sets on $L=0,1$, as we will
see in the following sections.

The system (\ref{dynsys}) is invariant under the discrete
transformation $(Q_0,Q_+,\tau) \rightarrow -(Q_0,Q_+,\tau)$, which
is a consequence of the invariance of the field equations under
time reversal. This reflects itself in the fact that the fixed
points of the system appear in pairs, except for the
``Einstein point" which is invariant
since $Q_0=Q_+=0$, see below.

The system~(\ref{dynsys}) admits the integral
\be\lb{genint} {\ce}(Q_0,Q_+,M_1,L) := \frac{3\,
(1-Q_0^2)^{4/3}\,V(L)}{4^{4/3}\,\Omega_D\, M_1^{1/3}} = E\, ,\ee
where $E$ is a constant ``energy.'' This conserved quantity is a
generalization of an integral that describes the solution
structure in the FRW case, cf.~Sec.~\ref{FRWSec}.
In~(\ref{genint}), the ``potential'' $V$ is defined as
\be\lb{potential} V := -  {\tilde \rho}\,x^2\: , \ee
where $\tilde{\rho}(x):=\rho/\rho_0$, and hence $V<0$ since
$\rho>0$; $V$ can be viewed as function of $x$ and thus of $L$ via
$x(L)$, and
\begin{equation}\label{dVdL}
\begin{array}{lll}
d V/d L & = & -(1+3 w) V [f(L) L(1-L)]^{-1}\quad\text{and}\,  \\[0.4ex]
V^\prime & = & -(1+3 w) V Q_0\, ,
\end{array}
\end{equation}
follows from $d\tilde{\rho}/dx = -3\tilde{\rho}(1+w)x^{-1}$.
Eq.~(\ref{dVdL}) in combination with~(\ref{dynsys}) leads to the
claimed conservation of $\ce$; $\ce^\prime=0$. The integral $\ce$
foliates the interior of the 4-dimensional state space
${}^4\textbf{X}$ into 3-dimensional hypersurfaces.

As an example of a potential $V$, consider again the combination
of two fluids with $w_0>w_1$; the choice
$\lambda(x)=(\rho_{10}/\rho_{00}) x^{3(w_0-w_1)}$ leads to
\begin{equation}\label{VL2}
V(L)  =  -C^2\,\left[\,L^{-(1+3w_0)}\,(1-L)^{1+3w_1}\,
\right]^{1/[3(w_0-w_1)]}\, ,
\end{equation}
with $C  =  C(\rho_{00},\rho_{10}) = const$. The special case of a
cosmological constant and a perfect fluid $p=w_0 \rho$ results in
\be\lb{VL} V(L) = \left[\frac{L(1-\cl)}{\cl
(1-L)}\right]^{2/[3(1+w_0)]}\,\left(\frac{\cl}{L}\right)\,V_{max}
\, ,\ee
where $V_{max}=V(\cl)$;  $\cl$ is defined by
$w(\cl)=-\textfrac{1}{3}$, which in the present case yields
$\cl:=(w_0+1/3)/(1+w_0)$.

To obtain a better feeling for our formalism we begin with a
description of the FRW cosmologies; first in the potential
approach, presumably familiar to a general reader, and then in the
dynamical systems picture.

\section{FRW cosmologies}
\label{FRWSec}

\subsection{The potential approach}
\label{FRWpotapp}

FRW models are characterized by the scale factor $\ell(t)$ subject
to the three key equations (see, e.g.,~\cite{book:gravitation})
\be \label{FRW} \dot{\ell}^{\,2}  =  \textfrac{1}{3}\rho \ell^2 - k\,
, \,\: {\ddot \ell} =  -\textfrac{1}{6}(\rho + 3p) \ell\, , \,\:
{\dot \rho} = -3(\rho + p) \, \dot{\ell}/\ell\, , \ee
where $t$ denotes the clock time along the fluid congruence which
is orthogonal to the symmetry surfaces. The first equation is the
Friedmann equation, which is an integral of the second one,
Raychadhuri's equation; the third equation describes local
conservation of energy; $k=+1,0,-1$ determines the sign of the
spatial curvature.
The combinations $\rho+3p$ and $\rho + p$ can be regarded as an
active gravitational and inertial mass-density, respectively;
positive (negative) $\rho+3 p$, i.e. $w>-1/3$ ($w<-1/3$) when
$\rho>0$, implies deceleration (acceleration).

To obtain a dimensionless formulation, $\ell$ is rescaled w.r.t.~a
reference time $t_0$: $x = \ell/\ell_0$ (with $\ell_0=\ell(t_0)$;
see p.~744 of~\cite{book:gravitation}). A natural choice for a
dimensionless time variable is $H(t_0)\, t$, which measures time
in ``Hubble units"; however, for our purposes we find it more
convenient to introduce $t_\rho := \sqrt{\rho_0/3}\: t$, i.e., we
measure time in ``energy-density units." Expressed in $x$ and
$t_\rho$, (\ref{FRW}) takes the form
\be \label{Friedmannpot} {\dot x}^2 = E - V\, ,\,\,{\ddot x} =
\textfrac{1}{2}(1+3w)V x^{-1}\, , \,\,
\dot{\tilde{\rho}} = -3 \tilde{\rho} (1+w)\, \dot{x}/x\:,
\end{equation}
where the overdot now denotes differentiation w.r.t.~$t_\rho$. The
equation of state enters via the functions $w(x)$ and
$V(x)$, cf.~(\ref{potential}); note that the
potential $V$ is scaled so that $V(1)=-1$. The first equation
in~(\ref{Friedmannpot}) can be interpreted as an ``energy"
integral
\begin{equation}\label{FRWce}
\ce = {\dot x}^2 + V = -(\rho - 3 H^2) \ell^2/(\rho_0 \ell_0^2)
=E\:,
\end{equation}
where the constant $E$ is given by $E = -(3 k)/(\rho_0 \ell_0^2)$.
The problem is thus analogous to the problem of a particle with
energy $E$ that moves in a potential $V(x)$. The condition $V<0$
implies that the open ($E>0$) and flat models ($E=0$) are either
forever expanding or contracting; however, the qualitative
behavior of closed models ($E<0$) depends on both the shape of the
potential and the value of $E$; we now focus on this case.

In addition to a non-negative energy-density, we henceforth also
assume a non-negative inertial mass-density $\rho + p$
(weak energy condition), i.e.,
$w\geq-1$ when $\rho>0$ (recall that $w = -1$ corresponds to a
cosmological constant). For simplicity we also assume that there
exists at most one value of $x$ such that $w(x) = -1/3$; hence we
require $dw/dx\, |_{w^{-1}(-1/3)}<0$, or equivalently $dw/d\rho\,
|_{w^{-1}(-1/3)}>0$. Under these assumptions $V$ has a maximum at
$x=w^{-1}(-1/3)$, since $dV/dx = -(1+3 w) V x^{-1}=0$ at $x=w^{-1}(-1/3)$ and $d^2 V/d
x^2 \,|_{w^{-1}(-1/3)} = -6 \tilde{\rho}^2
dw/d\tilde{\rho}\,|_{w=-1/3} < 0$.

\begin{figure}[h]
\centering
        \includegraphics[height=0.25\textwidth]{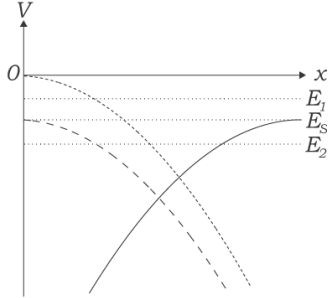}
        \caption{Qualitative shapes of potentials for FRW in the Cases
        (iii)[small dashed line], (iv)[solid line] and (v)[large dashed line]
        of Table~\ref{tabclosedFRWM}.}
    \label{figcaseIII_IV_V}
\end{figure}

\begin{figure*}[tc]
\centering
        \subfigure[Case (i)]{
        \label{figcaseI}
        \includegraphics[width=0.23\textwidth]{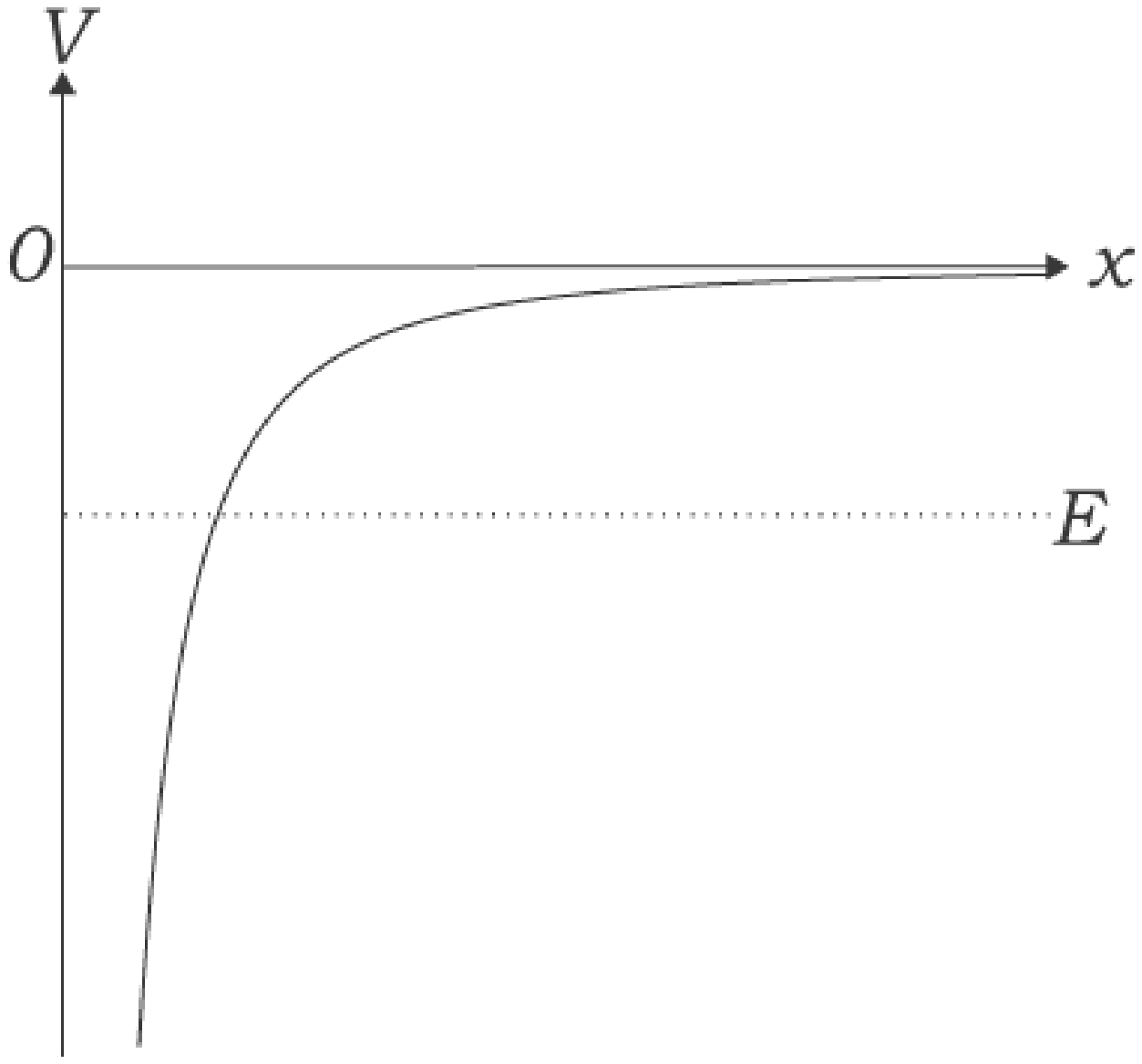}}\qquad\qquad\qquad
        \subfigure[Case (i). Phase portrait.]{
        \label{FRWsub0}
        \includegraphics[height=0.25\textwidth]{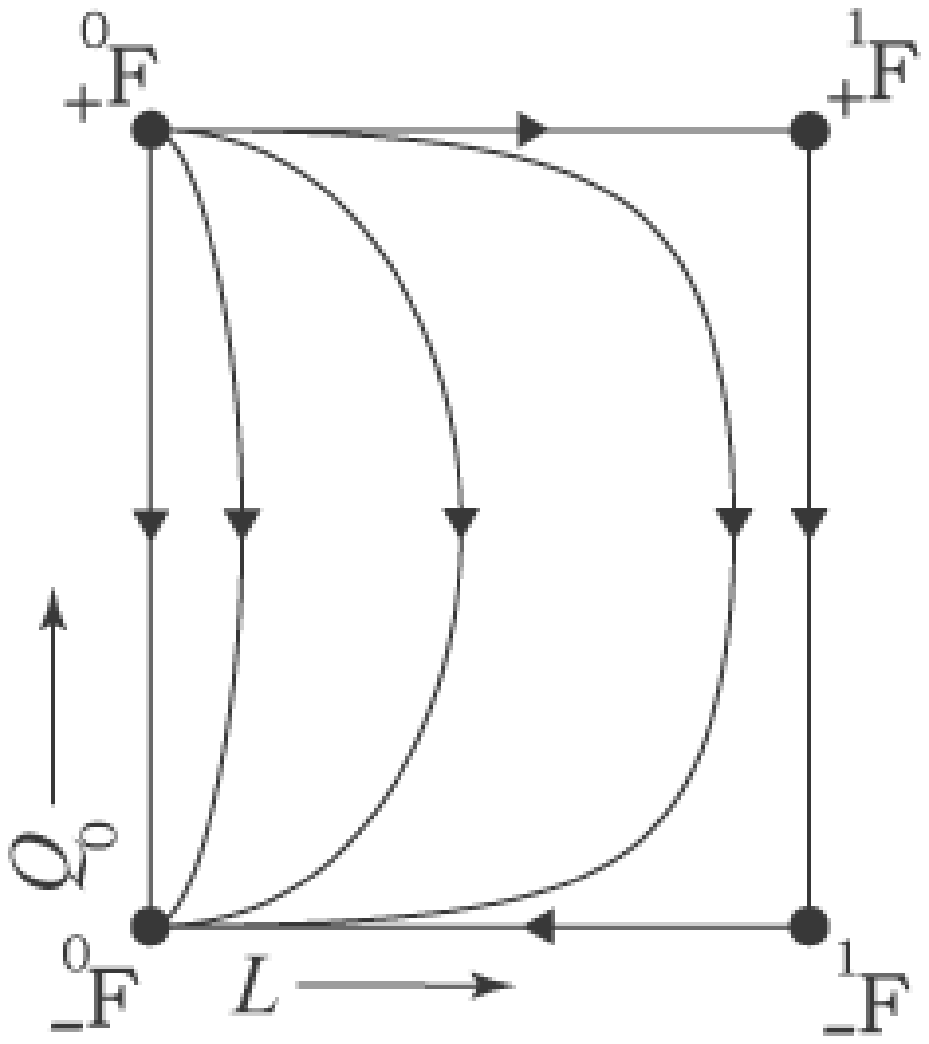}}\\
 \subfigure[Case (ii)]{
        \label{figcaseII}
        \includegraphics[width=0.23\textwidth]{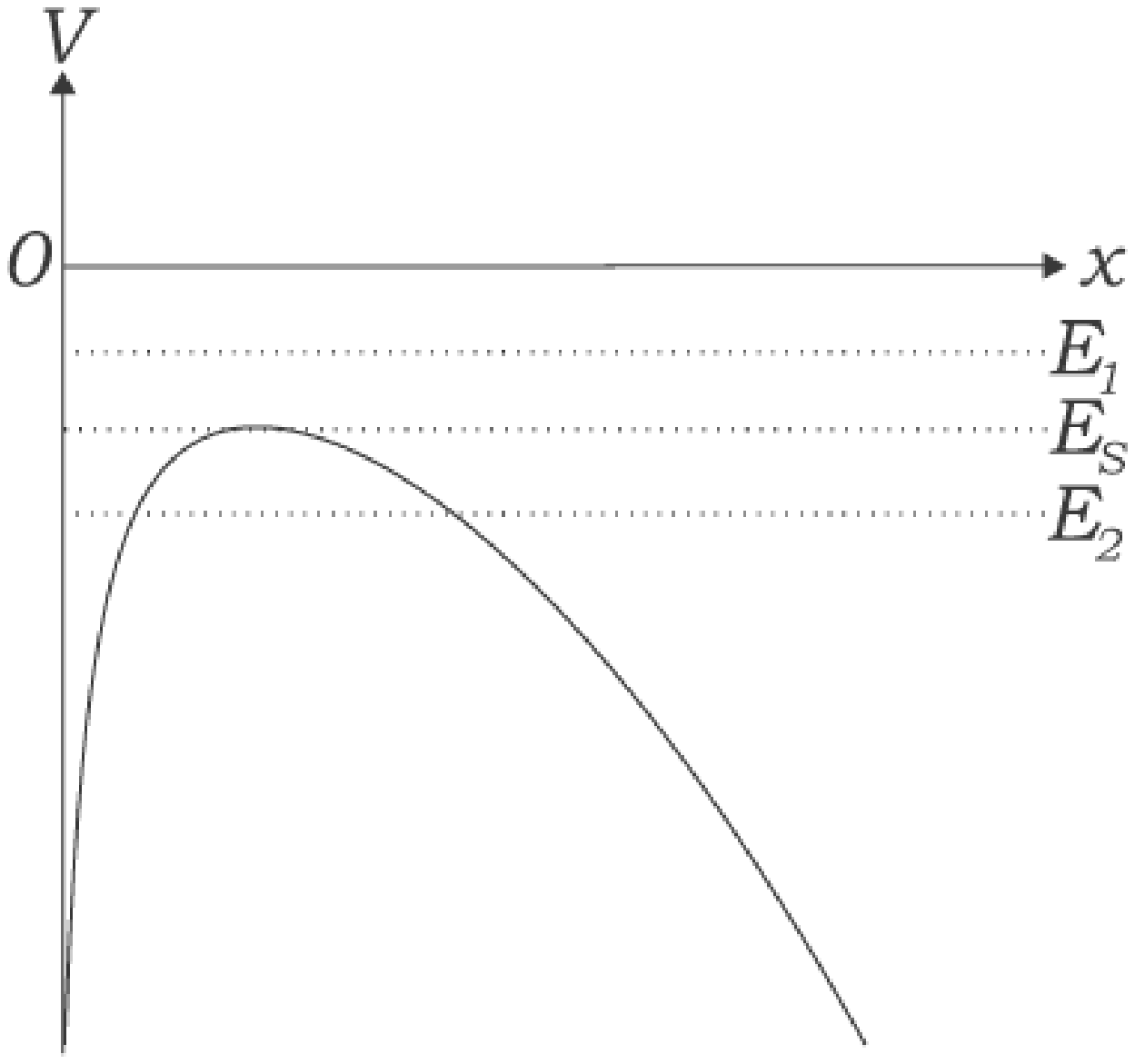}}\qquad\qquad\qquad
 \subfigure[Case (ii). Phase portrait.]{
        \label{FRWsub}
        \includegraphics[width=0.21\textwidth]{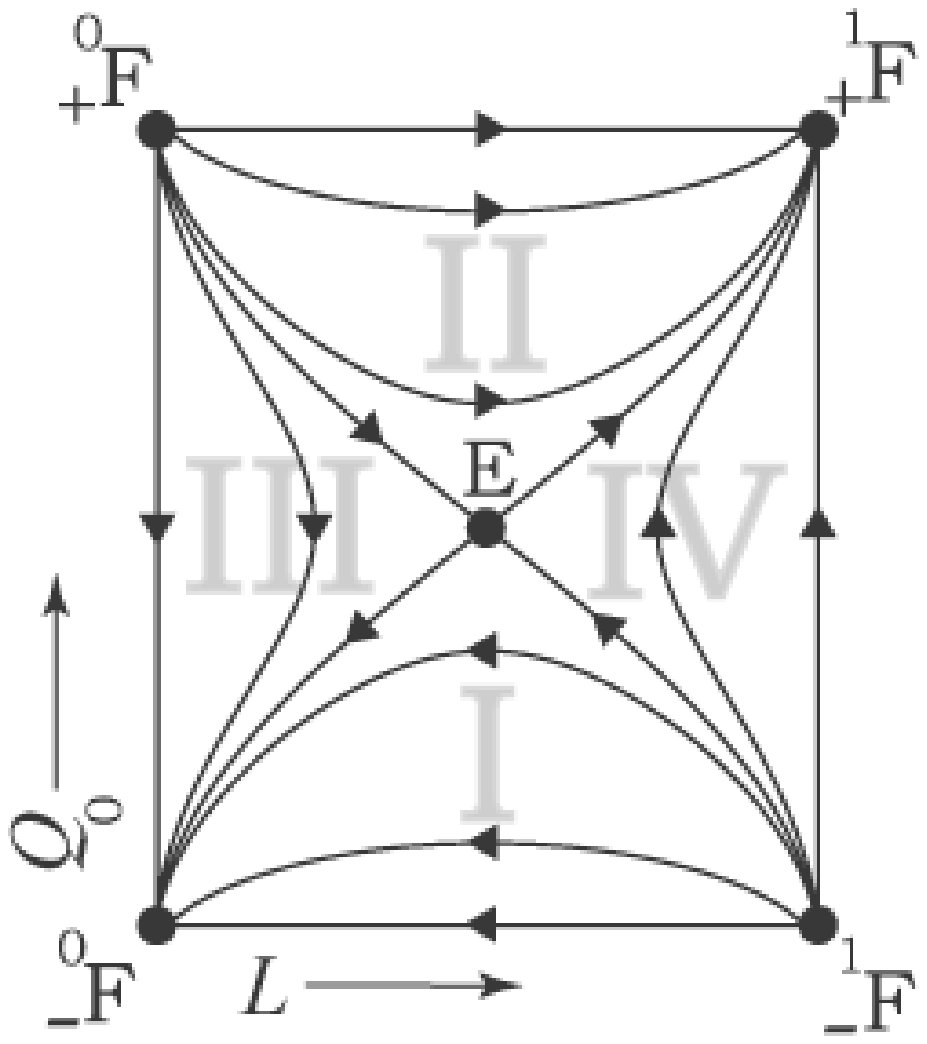}}
    \caption{Qualitative shapes of potentials for FRW Cases (i) and (ii)
             in Table~\ref{tabclosedFRWM}, with associated flows of
             the FRW dynamical system~(\ref{DFRW}).}
    \label{fig:potdia}
\end{figure*}

We distinguish five different cases; different matter assumptions
yield qualitatively different dynamical properties for the
associated FRW models, see Table~\ref{tabclosedFRWM}; the
corresponding potentials are depicted in
Figs.~\ref{figcaseIII_IV_V},~\ref{figcaseI}, and~\ref{figcaseII}.
\begin{table}[ht]
\begin{center}


  \begin{tabular}{c|c|l}

      &  &    \\[-0.3cm]
      Case & $w(x)$ & Qualitative FRW Dynamics  \\  \hline
    &  &    \\[-0.2cm]

     (i)  & $w>-1/3$, $\inf w > -1/3$& expanding--contracting  \\ \hline
       &  &    \\[-0.1cm]

     (ii) & $w>-1/3$ for small $x$ & $E=E_1$ : forever expanding/contracting \\
       & $w=-1/3$ for a unique $x$ & $E=E_s$ : expanding/contracting--Einstein static \\
       &   $w<-1/3$ for large $x$  & $E=E_2$ : expanding--contracting or reverse \\ \hline
       &  &    \\[-0.2cm]

     (iii)  & $w<-1/3$, $\sup w < -1/3$& contracting--expanding  \\ \hline
       &  &    \\

     (iv) & $w>-1/3$ for $x\rightarrow 0$ & $E=E_1$ : forever expanding/contracting \\
       & and $\lim_{x\rightarrow\infty}w=-1/3$ & $E=E_s$ : expanding--Einstein static \\
       &  & $E=E_2$ : expanding--contracting \\ \hline
       &  &    \\[-0.1cm]

      (v) & $w<-1/3$ for $x\rightarrow \infty$ & $E=E_1$ : forever expanding/contracting \\
       & and $\lim_{x\rightarrow 0}w=-1/3$ & $E=E_s$ : contracting--Einstein static \\
       &  & $E=E_2$ : contracting--expanding \\

 \end{tabular}
    \caption{Matter cases and the qualitative evolutionary
             behavior for closed FRW models.} \label{tabclosedFRWM}
 \end{center}

\end{table}

Figs.~\ref{figcaseIII_IV_V},~\ref{figcaseI}, and~\ref{figcaseII}
provide a good qualitative understanding of the FRW dynamics,
described in Table~\ref{tabclosedFRWM}, however, the exact
asymptotic behavior of solutions for $x\rightarrow 0, \infty$
depends on the asymptotic properties of the potential.
Asymptotically linear equations of state constitute simple
examples, since $w\rightarrow w_0$ for $x\rightarrow 0$ and
$w\rightarrow w_1$ for $x\rightarrow\infty$. Consider, e.g., a
source that consists of non-interacting perfect fluids with linear
equations of state; then $V(x) = -\tilde{\rho}x^2=
-\sum_i\tilde{\rho}_{i0}x^{-(1+3w_i)}$, where
$\tilde{\rho}_{i0}:={\rho}_{i0}/{\rho}_{0}$ and
$\sum_i\tilde{\rho}_{i0}=1$, and hence $\lim_{x\rightarrow 0} w =
\max w_i$ and $\lim_{x\rightarrow\infty} w = \min w_i$.

\subsection{The dynamical system approach}
\label{FRWdynsys}

In the dynamical systems picture the FRW models are represented by
solutions of the system~(\ref{dynsys}) on a particular invariant
manifold obtained by setting $Q_+=0$ and $Q_+^\prime=0$, which
implies that $M_1= \frac{1}{4}(1-Q_0^2)$ ($4 M_1^\prime =
(1-Q_0^2)^\prime$ follows). On the FRW subset the general system
reduces to
\begin{equation}\label{DFRW}
\begin{array}{lll}
Q_0' & = & -\textfrac{1}{8}(1-Q_0^2)(3+Q_0^2)(1+3w(L))\,  ,\\[0.7ex]
L' & = & f(L)L(1-L)Q_0\, .
\end{array}
\end{equation}
The conserved energy integral~(\ref{genint}) reads
\be\lb{FRWint} \ce(Q_0,L) = \frac{3(1-Q_0^2)}{3+Q_0^2}\,V(L)= E <
0\: ,\ee
cf.~also~(\ref{FRWce}). The conserved quantity determines the
orbits of~(\ref{DFRW}) in the FRW state space $\{(Q_0,L)\}$. To
simplify the subsequent discussion, we assume an asymptotically
linear equation of state that also strictly satisfies asymptotic
causality, i.e., $w_0, w_1 <1$.

Case (i), i.e., $w>-1/3$ everywhere and asymptotically (strong energy condition),
cf.~Table~\ref{tabclosedFRWM}, yields Fig.~\ref{figcaseI} in the
potential approach, while in the dynamical systems approach we
obtain the flow depicted in Fig.~\ref{FRWsub0}.
It is easy to see that the models initially expand, reach a point
of maximum expansion when $Q_0=0$, and then recollapse.

For Case (ii), i.e., matter domination in the limit $L\rightarrow
0$, ($-\frac{1}{3} < w_0 < 1$) and dark energy domination in the
limit $L\rightarrow 1$ ($-1\leq w_1 < -\frac{1}{3}$), the
potential has the form depicted in Fig.~\ref{figcaseII}. The
associated phase portrait in the dynamical systems approach is
given in Fig.~\ref{FRWsub}; the structure of the phase space
is as follows:
\begin{enumerate}
\item The energy $E_S=V_{max}$, where $V_{max}$ is the
maximum of the potential, yields the fixed point $\text{E}$ and
the four heteroclinic separatrix orbits, which originate from the
fixed points ${^{1}_+}\text{F}$, ${^{0}_{-}}\text{F}$ (or
$\text{E}$), and end at $\text{E}$ (or ${^{0}_-}\text{F}$,
${^{1}_{+}}\text{F}$). The point $\text{E}$ represents Einstein's
static universe; its position is $(0,\cl)$
where $\cl$ is determined by $w(\cl)=-1/3$;
${^{0,1}_\pm}\text{F}$ represent flat fluid FRW
models with linear equation of state (when $w_1=-1$,
${^1_\pm}\text{F}$ reduce to the de Sitter model). The
separatrix orbits divide the phase space into four regions that
contain models with distinct qualitative behavior.
\item $E = E_1>V_{max}$ yields regions I and II; region I (II)
contains forever contracting (expanding) models.
\item $E= E_2 < V_{max}$ yields regions III and IV. The orbits in
region III (IV) represent models that initially expand (contract),
reach a point of maximal (minimal) extension at $Q_0=0$, and then
contract (expand).
\end{enumerate}

The equations in~(\ref{dynsys}) and~(\ref{DFRW}) are completely
regular, this is in stark contrast to the non-regularity of the equations
used in~\cite{Oliveira/etal:1997,Barguine/etal:2001,olietal2002}.
Non-regularity leads to a misleading picture, e.g., compare
Fig.~\ref{FRWsub} with Fig.1 in~\cite{olietal2002}, where the
whole $L=0$ subset has been crushed to a point, which in turn
leads to a deformation of the interior orbit structure; moreover,
in contrast to $L=0$ in our approach, that point is not
part of the state space in~\cite{olietal2002}, since the
non-regularized equations blow up there.

Of the cases not discussed above, Case (iii) is closely related to Case (i);
in Cases (iv) and (v) the Einstein point lies on the boundaries $L=0$ and
$L=1$ respectively; although these cases can also be treated
easily, we refrain from doing so here.

Thus both the potential and the dynamical systems approaches
provide comprehensive information about the qualititative
structure of the FRW solution space, however, in contrast to the
potential approach the dynamical systems approach can be applied
to more general problems.

\section{LRS type IX cosmologies}
\label{LRSIX}

The foundation of our analysis of the LRS type IX models is the
regular dynamical system~(\ref{dynsys}), supplemented by the
conserved energy~(\ref{genint}). For simplicity we assume $w<1$
and, initially, an asymptotically linear equation of state. We
focus on Case (i), where $-1/3 < w_0, w_1 <1$, and Case (ii),
where $-1/3<w_0<1$ and $-1\leq w_1<-1/3$.

We begin with a local dynamical systems analysis: the fixed points
and the associated eigenvalues are given in
Table~\ref{tab4Dfixpoints}. All fixed points correspond to either
vacuum solutions or solutions that can be interpreted as perfect
fluid solutions with a linear equation of state: $\text{F}$
represents flat FRW solutions for a fluid with a linear equation
of state; Q stands for the LRS Kasner solution and T for the Taub
representation of Minkowski spacetime
(see~\cite{book:waiell97}); CS stands for the Bianchi type II
Collins and Stewart solution; X stands for a solution discussed,
e.g., in \cite{golell1999}; finally $\text{E}$ represents
Einstein's static universe. The upper left index refers to $L=0$
or $L=1$; the lower left index refers to $Q_0=+1$ or $Q_0=-1$,
except for $_\pm$X where $\pm$ indicates the signs in the
expressions for $Q_0$ and $Q_+$; the lower right index refers to
$Q_+=+1$ or $Q_+=-1$. Note that the fixed points $\text{E}$ and
$_\pm$X do not exist in Case (i) while ${^1_\pm}$CS do not exist
in Case (ii). Table~\ref{tab4Dfixpoints} shows that in Case (i)
all fixed points are saddles except for the source
$_+^0\text{T}_-$ and the sink $_-^0\text{T}_+$. In Case (ii) all
fixed points are saddles except for $_+^0\text{T}_-$ and
$_-^1\text{F}$, which are sources, and $_-^0\text{T}_+$ and
$_+^1\text{F}$, which are sinks.

\begin{table}[ht]
  \begin{center}
    \begin{tabular}{c|cccc|c}

      F.P & $Q_0$ & $Q_+$ & $M_1$ & $L$ & $\lambda_1$, $\lambda_2$, $\lambda_3$, $\lambda_4$ \\  \hline
      \\[-0.2cm]
      ${^A_\pm}$F         & $\pm 1$ & 0 & $ 0 $ & A
      & $\pm(1+3w_A)$, $\pm(1+3w_A)$, $\mp\frac{3}{2}(1-w_A)$, $\pm S(A)f_A$  \\
      ${^A_\pm}$Q$_{\pm}$ & $\pm 1$ & $\pm 1$ & 0 & A & $\pm 3(1-w_A)$, $\pm$2, $\mp$4, $\pm S(A)f_A$ \\
      ${^A_\pm}$T$_{\mp}$ & $\pm 1$ & $\mp 1$ & 0 & A & $\pm 3(1-w_A)$, $\pm$6, $\pm$12, $\pm S(A)f_A$ \\
      ${^A_\pm}$CS          & $\pm 1$ & $\pm\frac{(1+3w_A)}{8}$ & $\mathcal{W}_A$ & A
      & $\pm\frac{3}{4}(1+3w_A)$, $\mp\frac{6(1-w_A)}{8}\left(1\pm i\sqrt{c}\right)$,
                              $\pm S(A)f_A$ \\
      $_{\pm}$X           & $\pm\frac{2}{(1-3w_1)}$ & $\,\pm\frac{(1+3w_1)}{(1-3w_1)}$ & 0 & 1
      & $\mp\frac{6(1+3w_1)}{1-3w_1}$, $\mp\frac{3}{2}\frac{1-w_1}{1-3w_1}\left(1\pm\sqrt{d}\right)$,
                             $\mp\frac{2f_1}{1-3w_1}$ \\
      E                   & 0 & 0 & $\frac{1}{4}$ & $\cl$ & $\frac{3}{4}\sqrt{b}$, $-\frac{3}{4}\sqrt{b}$,
                             $i\sqrt{6}$, $-i\sqrt{6}$ \\

    \end{tabular}
  \end{center}
    \caption{Fixed points of the dynamical system~(\ref{dynsys}) and eigenvalues of the linearizations.
             Here $A$ takes the values $0$ or $1$; $\mathcal{W}_A :=3(1-w_A)(1+3w_A)/8^2$;
             $\cl$ is defined by $w(\cl):=-1/3$; $S(0)=1$, $S(1)=-1$, $b:=-2 w' f(\cl)\cl(1-\cl)$,
             $c:=(-3w_A^2+16w_A+3)/(2(1-w_A))$ and $d:=(24w_1^2+7w_1+1)/(1-w_1)$;
             $f(\cl)>0$ and $-w'=-dw/dL|_{\cl}>0$, which follows
             from $d\lambda/dx>0$ and $dw/dx|_{\cl}<0$.}
    \label{tab4Dfixpoints}
\end{table}

\subsection{The $\mathbf{w=const}$ case}

\begin{figure}[tc]

        \subfigure[$-1/3 < w=const < 1$]{
        \label{L0tent}
        \includegraphics[height=0.27\textwidth]{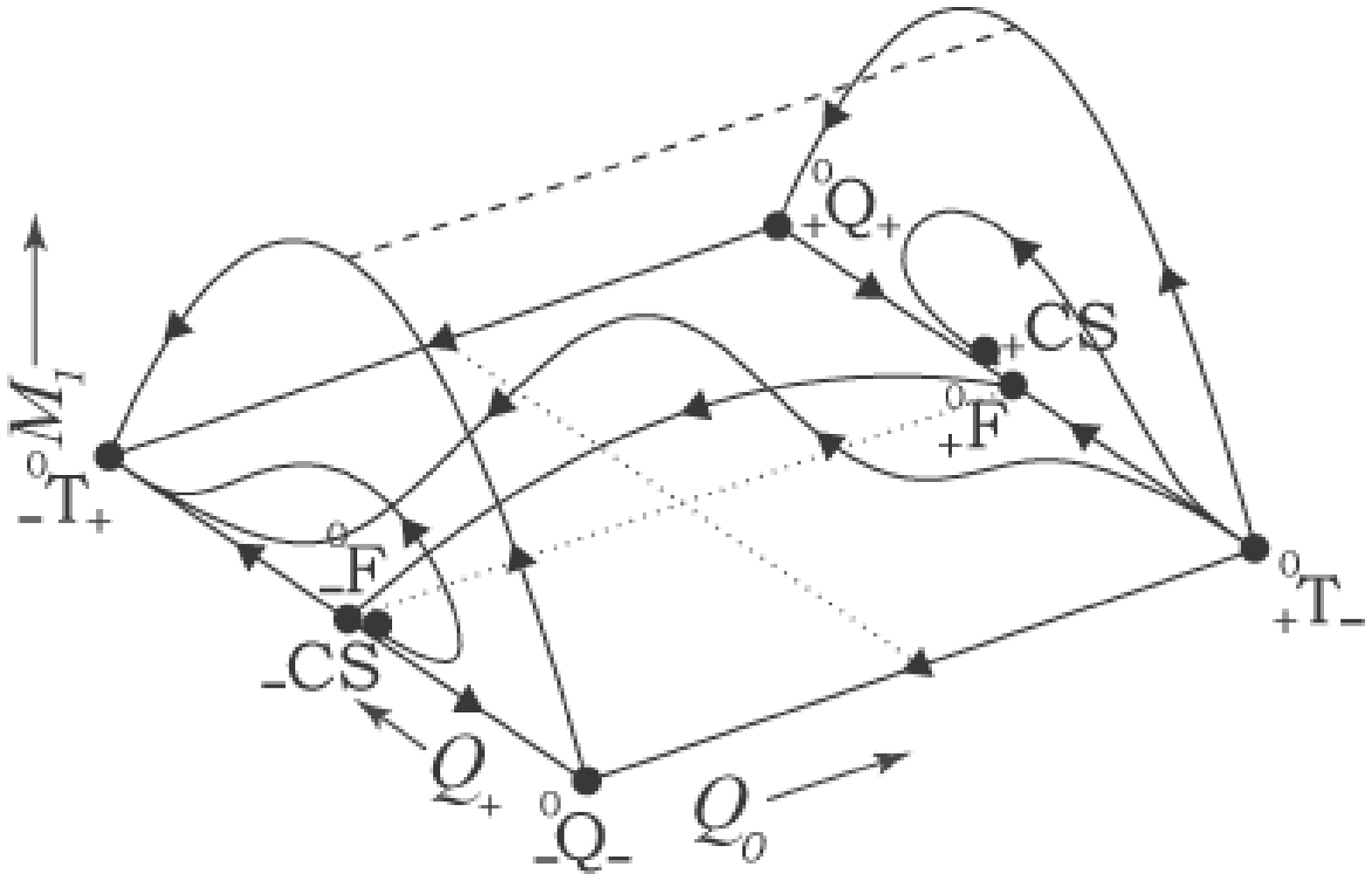}}\qquad
        \subfigure[$-1 \leq w=const < -1/3$]{
        \label{L1tent}
        \includegraphics[height=0.27\textwidth]{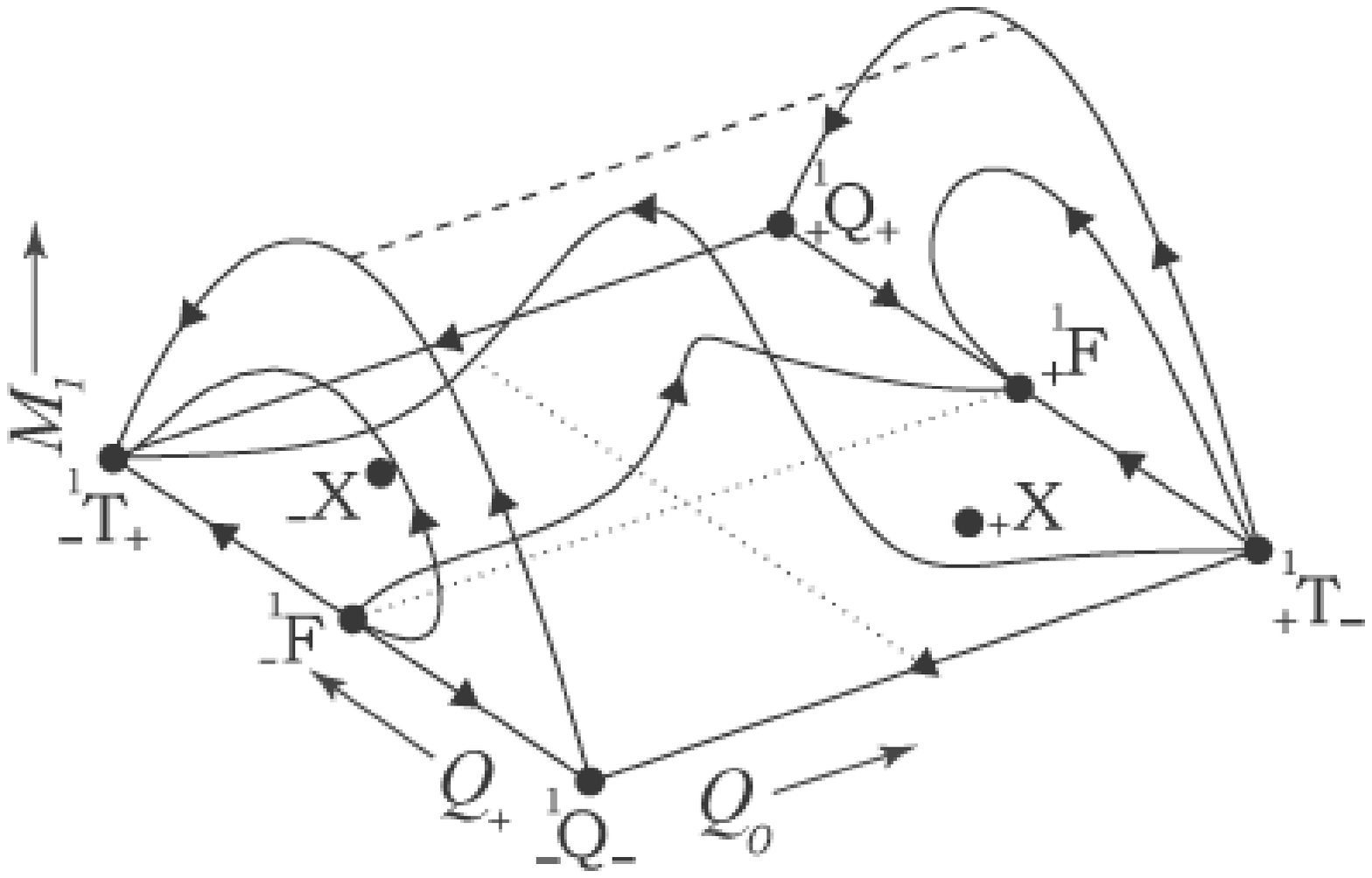}}\qquad
    \caption{Typical phase portraits for $-1/3 < w\equiv const < 1$ and $-1 \leq w\equiv const < -1/3$.}
    \label{Ltents}
\end{figure}

In the case of LRS type IX models with a linear equation of state
$w\equiv const$, the equation for $L$ decouples
from~(\ref{dynsys}); therefore it suffices to consider the state
space $\{(Q_0,Q_+,M_1)\}$. We distinguish two ranges for $w$ that
generate different sets of fixed points: $-1/3 < w < 1$ and $-1
\leq w  < -1/3$ (we refrain from giving the special case $w=-1/3$).
Note that the equation system (and thus the solution
structure) for $-1/3 < w < 1$ coincides with the system on the
subsets $L=0$ and $L=1$ in Case (i) and the subset $L=0$ in Case
(ii); $-1 \leq w < -1/3$ is identical to the $L=1$ subset in Case
(ii). The state spaces, and some orbits, are given in
Fig.~\ref{Ltents}.

\subsection{LRS type IX models with $\mathbf{-\frac{1}{3}< w<1}$}
\label{wbigger-1/3}

Next we consider Case (i), i.e., models with $-1/3<w<1$
and $-1/3< w_0, w_1 < 1$.
We prove that all such models expand from a singularity, reach a point of
maximum expansion, and subsequently recollapse to a singularity.
Hereby the $\omega$-limit ($\alpha$-limit) of every
orbit is located on the 2-plane
$(L=0) \cap (Q_0 = -1)$ ($(L=0) \cap (Q_0 = +1)$).

The subcase $0< w <1$ is covered by the general results
of~\cite{Lin/Wald:1990} and~\cite{Rendall:1995}, where it is
proved, for the general Bianchi type IX case, that models do not
expand forever, instead they re-collapse to a singularity where
certain curvature invariants and the density diverge. Our
dynamical systems approach yields more details about the
asymptotic dynamics and the methods of our proof are entirely
different from those in~\cite{Lin/Wald:1990,Rendall:1995}: our
starting point is the dynamical system~(\ref{dynsys}).

On the hypersurface $Q_0 =0$ we obtain $F>0$ and therefore
$Q_0^\prime <0$; hence, $Q_0 = 0$ acts as a ``semipermeable
membrane'' for the flow of the system~(\ref{dynsys}). The
hypersurface cuts the interior state space in two halves, a future invariant%
     \footnote{In Appendix~\ref{nomenclature} we give a brief introduction
     to dynamical systems nomenclature.}
half $Q_0 < 0$, where $L$ monotonically decreases, i.e., $L^\prime
< 0$, and a past invariant half $Q_0>0$, where $L^\prime >0$. It
follows that fixed points and periodic orbits are excluded in the
interior of the state space.

Application of the principle of monotone functions, see,
e.g.,~\cite{book:waiell97}, and investigation of
the structure of the flow on $Q_0
=0$, yields that the $\omega$-limit of the future invariant set
$Q_0<0$ must be contained in $L=0$; analogously the $\alpha$-limit
of every orbit in $Q_0>0$ must lie on $L=0$. More specifically,
the $\omega$-limit is located on the 2-plane $(L=0) \cap (Q_0 =
-1)$. To show this, consider an orbit $(Q_0,Q_+,M_1,L)(\tau)$ in
$Q_0< 0$. Since $L\rightarrow 0$ for $\tau\rightarrow \infty$, the
function $V$, cf.~Fig.~\ref{figcaseI}, satisfies $V\rightarrow
-\infty$. The integral of motion~(\ref{genint}) fulfills
\begin{equation}\label{limitof1-Q02}
(1-Q_0^2)^{4/3} = \frac{4^{4/3}\,\Omega_D\, M_1^{1/3}\,E}{3\,V}
\:\rightarrow\: 0 \qquad (\tau\rightarrow\infty)\:,
\end{equation}
and thus $Q_0 \rightarrow -1$ ($\tau\rightarrow\infty$), as
claimed; analogously, the $\alpha$-limit of $Q_0>0$ lies on $(L=0)
\cap (Q_0 = +1)$.

As regards the future asymptotic behavior of orbits in $Q_0>0$
(and the $\alpha$-limit of $Q_0<0$), there exist, a priori, two
possibilities: an orbit either passes through $Q_0=0$ into
$Q_0<0$, and thus converges to $(L=0) \cap (Q_0 = -1)$, or it
remains in $Q_0>0$ for all times. However, a proof by
contradiction shows that the second scenario is excluded: let us
assume that the solution satisfies $Q_0>0 \:\forall\, \tau$. The
monotonicity principle then implies $L\rightarrow 1$ for
$\tau\rightarrow \infty$. Hence, $V\rightarrow 0$ and,
from~(\ref{genint}),
\begin{equation}
\Omega_D M_1^{1/3} = \frac{3\,(1-Q_0^2)^{4/3}\,V}{4^{4/3}\,E}
\:\rightarrow\: 0 \qquad (\tau\rightarrow\infty)\:.
\end{equation}
It follows that either $\Omega_D\rightarrow 0$ or $M_1\rightarrow
0$ as $\tau\rightarrow \infty$. This restricts the $\omega$-limit
of the orbit to the 2-surface

\begin{equation}\label{noomegasurface}
\left[(L=1)\cap(Q_0\geq 0)\cap(\Omega_D=0)\right]\:\, \cup
\left[(L=1)\cap(Q_0\geq 0)\cap (M_1=0)\right]\:.
\end{equation}
Since the $\omega$-limit set of the orbit is located on a
2-surface, the $\omega$-limit is either a fixed point, a periodic
orbit, a heteroclinic cycle, or it is the whole 2-surface that
acts as the attractor. The local analysis of ${}^1_+\text{Q}_{+}$,
${}^1_+\text{T}_{-}$, and ${}^1_+\text{F}$ reveals that these
fixed points are saddles such that no orbit from the interior of
the state space can be attracted. Moreover, the
surface~(\ref{noomegasurface}) does neither contain periodic
orbits nor heteroclinic cycles; the possibility that the
$\omega$-limit of the orbit is the whole 2-surface is excluded by
the structure of the flow as well. Since the
2-surface~(\ref{noomegasurface}) does not act as an $\omega$-limit
set, we have arrived at a contradiction; $Q_0>0 \:\forall\, \tau$
is thus excluded and therefore the $\omega$-limit of an orbit in
$Q_0>0$ lies on $(L=0) \cap (Q_0 = -1)$.

Note that the ``saddle structure" of the
surface~(\ref{noomegasurface}) is valid for any value $w_1>-1/3$;
it follows that it is not necessary to assume asymptotic
linearity of the equation of state, i.e., $w\rightarrow w_1 > -1/3$
is not required, $(\inf w) > -1/3$ is a sufficient assumption.

\subsection{LRS type IX models with $\mathbf{-1 < w < 1}$}
\label{w-1--1/3}

Consider now Case (ii), i.e., models with $-1/3< w <1$ for
$L<\cl$, $w(\cl)=-1/3$, $-1\leq w< -1/3$ for $L>\cl$, and
$w^\prime := dw/dL\,|_{\cl} < 0$,%
    \footnote{For brevity, a prime also denotes differentiation
      w.r.t.\ the argument $L$ for functions like $w(L)$ and $V(L)$.}
whereby $V$ has a single maximum, cf.~Sec.~\ref{FRWpotapp}.

Below we give a comprehensive description of the dynamics of
models in a neighborhood of the Einstein point and of the
asymptotic behavior of these models. The analogous problem has
been investigated in~\cite{Oliveira/etal:1997,Barguine/etal:2001}
for the special case of a fluid with a linear equation of state
and a cosmological constant, however, we arrive at different
conclusions. We demonstrate that the asymptotic behavior of models
close to the Einstein point is predictable and not governed by
``homoclinic phenomena and chaos''. In addition we also show
several global results.

Our analysis is based on our dynamical systems formulation,
i.e., the regular system~(\ref{dynsys}) on the compact state space ${}^4\textbf{X}$;
the analysis rests on three cornerstones: the local
dynamical systems analysis of the Einstein point $\text{E}$, in
particular center manifold reduction; the use of the conserved
quantity $\ce$; the recognition and use of the close relationship
between the dynamics of the full system and the FRW subset.

The {\em first cornerstone\/} in our analysis is the local
dynamical systems analysis of the Einstein point E. Two
eigenvalues, $\lambda_{1,2} = \pm3\sqrt{b}/4$, are associated with
eigenvectors that lie in the FRW plane,
$\mathbf{v}_{1,2}=(1,0,0,\pm 2 \sqrt{b}/(3 w^\prime))$,
where $b:=-2w^\prime f(\cl)\cl(1-\cl)$;
two purely imaginary eigenvalues, $\lambda_{3,4} = \pm i
\sqrt{6}$, lead to a rotation in the plane orthogonal to the FRW
plane. Hence the linear space at $\text{E}$ decomposes into the
direct sum $E^s\oplus E^u\oplus E^c$ of a one-dimensional stable
(unstable) subspace $E^s$ ($E^u$), and a two-dimensional center
subspace $E^c =\langle (0,1,0,0), (0,0,1,0)\rangle$.

To adapt to $\text{E}$ and the structure $E^s\oplus E^u\oplus E^c$
we introduce new variables,
\be\label{adaptedcoords} \tilde{q}_0    =   \frac{1}{2} \left(Q_0
+ \frac{9\, w^\prime}{8\, \lambda_1} \:(L-\cl)\, \right)\, ,\:\:
q_+ = Q_+\, ,
 m_1  = M_1-\frac{1}{4}\, ,\:\: \tilde{l} = \frac{1}{2}
\left(Q_0 - \frac{9\,w^\prime}{8\,\lambda_1} \:(L-\cl)\, \right)\,
, \ee
which transform the dynamical system~(\ref{dynsys}) to
\begin{equation}\label{fullschematically}
\begin{pmatrix}
\tilde{q}_0^\prime \\
\tilde{l}^\prime
\end{pmatrix}  =
\begin{pmatrix}
\lambda_1 & 0 \\
0 &  \lambda_2
\end{pmatrix}
\begin{pmatrix}
\tilde{q}_0 \\
\tilde{l}
\end{pmatrix}
+
\begin{pmatrix}
N_{{\tilde q}_0} \\
N_{{\tilde l}}
\end{pmatrix}\,,\quad
 \begin{pmatrix}
q_+^\prime \\
m_1^\prime
\end{pmatrix} =
\begin{pmatrix}
0 & 4 \\
-\frac{3}{2} & 0
\end{pmatrix}
\begin{pmatrix}
q_+ \\
m_1
\end{pmatrix}
+
\begin{pmatrix}
N_{q_+} \\
N_{m_1}
\end{pmatrix}\, ,
\end{equation}
where the nonlinear terms are collected in the respective terms $N
= N(\tilde{q}_0,q_+,m_1,\tilde{l})$.

The invariant center manifold $M_E^c$ is two-dimensional,
tangential to $E^c$, and contains $\text{E}$. Locally it is
represented by the graph of a function $h: E^c\rightarrow
E^s\oplus E^u$,
\begin{equation}\label{graphh}
 (q_+,m_1) \mapsto (\tilde{q}_0,\tilde{l})=(h_{q_0}(q_+,m_1),
h_{l}(q_+,m_1))\, ,
\end{equation}
which satisfies $h(0) = 0$ and $\text{grad} \:h(0) =0$. Center
manifold reduction, see, e.g.~\cite{cra1991}, reduces the full
nonlinear system~(\ref{fullschematically}) to the locally
\textit{equivalent} system
\begin{equation}\label{reducedschematically}
\begin{pmatrix}
\tilde{q}_0^\prime \\
\tilde{l}^\prime
\end{pmatrix}  =
\begin{pmatrix}
\lambda_1 & 0 \\
0 &  \lambda_2
\end{pmatrix}
\begin{pmatrix}
\tilde{q}_0 \\
\tilde{l}
\end{pmatrix}\:, \quad
 \begin{pmatrix}
q_+^\prime \\
m_1^\prime
\end{pmatrix}  =
\begin{pmatrix}
0 & 4 \\
-\frac{3}{2} & 0
\end{pmatrix}
\begin{pmatrix}
q_+ \\
m_1
\end{pmatrix}
+
\begin{pmatrix}
N_{q_+} \\
N_{m_1}
\end{pmatrix}\ ,
\end{equation}
where $N= N((h_{q_0}(q_+,m_1),q_+,m_1,h_l(q_+,m_1))$.

In general it is impossible to obtain the graph of $h$ in terms of
elementary functions, however, approximate solutions can be
obtained by series expansions:
\begin{equation}\label{centermanifoldexpansion}
h_{q_0}(q_+,m_1)  = \tilde{q}_0 = \frac{1}{2} \left(q_0 +
\frac{9}{8}\,\frac{w^\prime}{\lambda_1} \, l \, \right)\, , \quad
h_{l}(q_+,m_1) = \tilde{l} = \frac{1}{2} \left(q_0  -
\frac{9}{8}\,\frac{w^\prime}{\lambda_1} \, l \, \right)\:,
\end{equation}
where
\begin{equation}\label{centermanifoldexpansionvar}
q_0  =  -2^4w^\prime\,k\, q_+ m_1 + O(\|(q_+,m_1)\|^3)\:,\quad l =
\textfrac{3 b+2^6}{3}k\, q_+^2 + \textfrac{2^{9}}{3^2}k\, m_1^2+
O(\|(q_+,m_1)\|^3)\, ,
\end{equation}
where $k:=-2^4/(3w^\prime(2^7+3b))$,
$b:=-2w^\prime\,f(\cl)\cl(1-\cl)$.

The coordinates in~(\ref{reducedschematically}) are \textit{not}
the coordinates~(\ref{adaptedcoords}), but coordinates adapted to
the center manifold structure. We assume that the two coordinate
systems agree to first order; this is justified since $E^u\oplus
E^s$ coincides with the FRW invariant subset and because
of~(\ref{centermanifoldexpansion}) and
(\ref{centermanifoldexpansionvar}). To avoid excessive notation we
use the same symbols for the coordinates of~(\ref{adaptedcoords})
and~(\ref{reducedschematically}).

The reduced system~(\ref{reducedschematically}) represents a
decoupled system, and hence the properties of the flow in the
neighborhood of $\text{E}$ can be deduced straightforwardly. The
center manifold $M_E^c$ is represented by $\tilde{q}_0=\tilde{l} =
0$, and the induced dynamical system by the $(q_+,m_1)$-system
of~(\ref{reducedschematically}). The solutions are ``circular''
periodic orbits centered about the fixed point $\text{E}$, as we
prove below. The fixed point $\text{E}$ is an $\omega$-limit for
two orbits and an $\alpha$-limit for another pair, which follows
from setting $q_+=m_1=0$ and $\tilde{q}_0=0$ (or $\tilde{l}=0$);
the four orbits are the separatrix orbits in the FRW plane, see
Fig.~\ref{FRWsub}. Each periodic orbit acts as $\omega$-limit for
two one-parameter families of orbits: in the invariant set
$\tilde{q}_0=0$, Eq.~(\ref{reducedschematically}) describes one
one-parameter family of orbits in the half-space $(\tilde{l}>0)$
(another one in $\tilde{l}<0$) that spiral down (up) towards the
periodic orbit at $\tilde{l}=0$; similarly, in the invariant space
$\tilde{l}=0$, the periodic orbit is the $\alpha$-limit for
another two one-parameter families of spirals. For general initial
data $(\tilde{q}_0\neq 0, \tilde{l}\neq 0)$ the closed orbits act
as a ``saddles'', see Fig.~\ref{spiral} below; we thus observe a
natural generalization of the FRW picture.

The {\em second cornerstone\/} in our analysis is the conserved
quantity $\ce$. To locate $\text{E}$ at the origin we introduce
the variables
\begin{equation}
q_0 = Q_0\, , \,\, q_+ = Q_+\, , \,\, m_1 = M_1-1/4\, , \,\, l =
L-\cl\, ,
\end{equation}
so that Eq.~(\ref{genint}) is transformed to
\begin{equation}\label{Einnew}
{\mathcal E}(q_0,q_+,m_1,l) :=
\frac{3\,(1-q_0^2)^{4/3}\,V(\cl+l)}{(3-4 q_+^2-4 m_1) (1+4
m_1)^{1/3}} = E\, .
\end{equation}
To discuss the 3-hypersurfaces of constant energy, $\ce=E$, in a
neighborhood of E, we expand~(\ref{Einnew}) up to second order,
using $V(\cl+l) = V_{max} + V^{\prime\prime}(\cl) l^2$ where
$V_{max} = V(\cl)$ and $V^{\prime\prime}<0$,
\begin{equation}\label{3hyp}
-4 \frac{V_{max}}{E}  q_0^2 + \frac{3}{2}
\frac{V^{\prime\prime}(\cl)}{E} \,l^2  +  \|(q_+,m_1)\|^2  = 3
\left(1-\frac{V_{max}}{E} \right)\, ,
\end{equation}
where $\|(q_+,m_1)\|^2:= 4q_+^2 + 16 m_1^2/3$.

We employ $\ce$ to prove that the orbits on the center manifold,
i.e., the solutions of the $(q_+,m_1)$-system
in~(\ref{reducedschematically}), are periodic. The idea is simple:
the intersection of a hypersurface $\ce=E$ with a transversal
surface that is itself invariant under the flow of the dynamical
system yields an orbit with energy $E$ on that surface. It is not
possible to study $M_E^c \cap ({\mathcal E}=E)$ directly, however,
the investigation of $E^c \cap ({\mathcal E}=E)$ turns out to be
sufficient. We therefore set $q_0 = l = 0$ in~(\ref{Einnew}), and
obtain $3-(3-4 q_+^2-4 m_1) (1 + 4 m_1)^{1/3} = 3 (1 -V_{max}/E)$,
which describes a family of closed curves, parametrized by $E$. In
an appropriately small neighborhood of the origin, these curves
can be approximated by $\|(q_+,m_1)\|^2 = 3 (1 - V_{max}/E) =
const$, where $E < V_{max}$, cf.~(\ref{3hyp}) with $q_0 = l = 0$.
Since $\text{grad}\, \ce \sim \big(0, c_1 q_+, c_2 (4 m_1 +
q_+^2),0\big)$, where $c_1$, $c_2$ are non-vanishing, it follows
that the surface ${\mathcal E}=E$ is orthogonal to $E^c$ at the
intersection; this guarantees the intersection to be a closed
curve also when $E^c$ is continuously deformed to $M_E^c$.

{\em Every\/} solution in a sufficiently small neighborhood of
$\text{E}$ obeys the relation $\|(q_+,m_1)\|^2 = const$
approximately, since the equations for $(q_+,m_1)$ decouple and
yield $\|(q_+,m_1)\|^2 = const$ in~(\ref{reducedschematically}).
Hence the hypersurface $\|(q_+,m_1)\| = const$ is a (locally)
invariant set. Each orbit in this invariant set gives rise to a
one-parameter family of orbits that differ only w.r.t.~the (polar)
angle along $\|(q_+,m_1)\| = const$. Therefore, one can represent
an orbit --modulo its phase-- by an orbit in the 3-space
${}^3\textbf{X} = \{\,\big(q_0,l, \|(q_+,m_1)\|\,\big)\,\}$ that
fulfills $\|(q_+,m_1)\| = const$; hereby the four-dimensional
dynamical picture is reduced to a three-dimensional one.

The state space ${}^3\textbf{X}$ is endowed with a Lorentzian
scalar product of signature ($-$$+$$+$), Eq.~(\ref{3hyp})
describes 2-surfaces of constant Lorentz norm. The light cone,
$E=V_{max}$, divides the state space into \textit{three}
disconnected regions: $E>V_{max}$ yields region I, the
chronological past of $\text{E}$, and region II, the chronological
future; $E<V_{max}$ yields region III/IV, the spacelike region,
see Fig.~\ref{regions}. Note that the disconnected regions III and
IV of the FRW picture~\ref{FRWsub} are merged to one connected
region III/IV. The coordinates $\tilde{q}_0$ and $\tilde{l}$ are
null coordinates w.r.t.~the metric; this can be seen
from~(\ref{adaptedcoords}) and the relation $V^{\prime\prime}(\cl)
= -3 w^\prime V_{max} \,[ f(\cl) \cl (1-\cl)^{-1}]^{-1}$, which
can be derived from~(\ref{dVdL}).

The {\em third cornerstone\/} in our analysis is the close
connection to the FRW case. In ${}^3\textbf{X}$, the FRW plane is
given by $\|(q_+,m_1)\|=0$. The separatrix orbits, $E=V_{\max}$,
are given by the null rays ${\tilde q}_0=0$ and $\tilde{l}=0$. Let
$\tilde{\nu}^+_q$ ($\tilde{\nu}^-_q$) denote the future (past)
null ray along the ${\tilde q}_0$-axis, and $\tilde{\nu}^\pm_l$
the null rays along the $\tilde{l}$-axis, see Fig.~\ref{nullcone}.
From Sec.~\ref{FRWdynsys} we know the global asymptotics,
\begin{equation}\label{nullraysglobal}
\begin{split}
&\tilde{\nu}^+_q  \leftrightarrow (\text{E}\rightarrow
{}^1_+\text{F})\, ,\qquad \tilde{\nu}^+_l  \leftrightarrow
({}^0_+\text{F}\rightarrow \text{E})\, ,\\
&\tilde{\nu}^-_q\leftrightarrow (\text{E}\rightarrow
{}^0_-\text{F})\, ,\qquad \tilde{\nu}^-_l \leftrightarrow
({}^1_-\text{F}\rightarrow \text{E})\:.
\end{split}
\end{equation}
Orbits close to $\tilde{\nu}^{\pm}_{q,l}$ are dragged along and
thus exhibit similar asymptotic behavior as $L\rightarrow 0,1$.
However, in the general picture the FRW attractor ${}^0_-\text{F}$
is generalized to $(L=0)\cap (Q_0 =-1)$, which
describes the contracting Bianchi type
II boundary with a linear equation of state characterized by $w_0$
(the FRW source ${}^0_+\text{F}$ is generalized to $(L=0)\cap (Q_0
=+1)$, which describes the expanding version of the type II
boundary); see App.~\ref{asystates}. A priori the FRW source
${}^1_-\text{F}$ becomes $(L=1)\cap (Q_0 =-1)$, however, on that
surface it is only the fixed point ${}^1_-\text{F}$ that can act
as an $\alpha$-limit; a similar $\omega$-limit statement holds for
${}^1_+\text{F}$.

A combination of the ideas of our analysis yields an effective
description of the dynamics close to $\text{E}$: in
${}^3\textbf{X}$, an orbit with a given energy $E$ lies on a
hyperboloid (or light cone) of constant Lorentz norm,
cf.~(\ref{3hyp}); simultaneously, every orbit satisfies
$\|(q_+,m_1)\|=const$. Therefore, the intersections of the
energy-hyperboloids with the planes $\|(q_+,m_1)\|=const$ yield a
complete description of all orbits in ${}^3\textbf{X}$;
furthermore, the asymptotics of~(\ref{nullraysglobal}) and the
above generalization determines the asymptotics of the orbits. We
investigate the scenarios case by case:

\begin{figure}[tc]
\centering
        \includegraphics[height=0.34\textwidth]{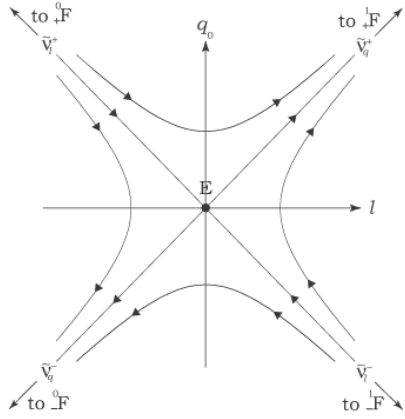}
    \caption{The FRW plane $\|(q_+,m_1)\|=0$ in the space ${}^3\textbf{X}$.
      Four null rays, $\nu^{\pm}_{q,l}$, converge to the Einstein point E
      as $\tau\rightarrow\pm\infty$.}
    \label{nullcone}
\end{figure}

\begin{figure}[htp]
\centering
        \includegraphics[width=0.34\textwidth]{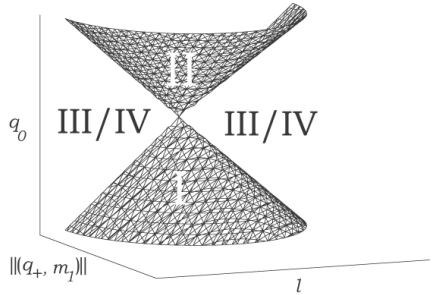}
    \caption{The light cone $E=V_{max}$ in ${}^3\textbf{X}$
      divides the state space in three regions that generalize
      the regions in the FRW picture~\ref{FRWsub}.}
    \label{regions}
\end{figure}

\begin{figure}[tc]
\centering
        \includegraphics[width=0.3\textwidth]{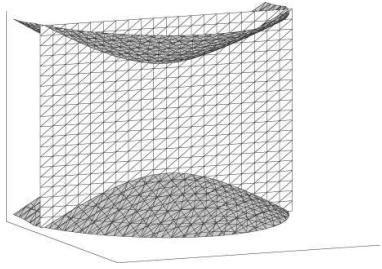}
    \caption{Case $E>V_{max}$.
      Intersection of the mass hyperboloids of energy $E$
      with a plane $\|(q_+,m_1)\|=const$ in ${}^3\textbf{X}$ yields two hyperbolas, one
      in region I and one in region II. The axes are the same as in Fig.~(\ref{regions}).}
    \label{spacehyps}
\end{figure}

When $E>V_{max}$, Eq.~(\ref{3hyp}) describes two ``mass''
hyperboloids in ${}^3\textbf{X}$; one in region I and one in
region II, see Fig.~\ref{spacehyps}. Intersection with planes
$\|(q_+,m_1)\|=const$ yields a family of spacelike hyperbolas
which are parallel to null rays asymptotically,
Fig.~\ref{spacehyps}; this indicates that the hyperbolas of region
I originate from ${}^1_-\text{F}$, as $\tilde{\nu}^-_l$ does, and
end on $(L=0)\cap (Q_0 =-1)$ like $\tilde{\nu}^-_q$. Analogously,
the hyperbolas in region II originate from $(L=0)\cap (Q_0 =+1)$
and end at ${}^1_+\text{F}$. The described global asymptotic
behavior of orbits also remains true for orbits remote from
$\text{E}$, as proved in App.~\ref{globaldynamicalfeatures}. Thus
the FRW picture is generalized: region I (II) contains models that
contract (expand) forever.

When $E=V_{max}$, Eq.~(\ref{3hyp}) describes the light cone in
${}^3\textbf{X}$, see Fig.~\ref{regions}. Intersection with the
FRW plane $\|(q_+,m_1)\|=0$ yields the four null rays discussed
above, cf.~(\ref{nullraysglobal}) and Fig.~\ref{nullcone}.
Intersection with planes $\|(q_+,m_1)\|=const>0$ yields a family
of hyperbolas that resemble those in the $E>V_{max}$ case; the
global behavior is thus analogous.

Finally, when $E<V_{max}$, Eq.~(\ref{3hyp}) describes one
hyperboloid of positive Lorentz norm in the region III/IV in
${}^3\textbf{X}$. We distinguish three subcases:

Case (a): Intersection with planes $\|(q_+,m_1)\|=const< 3 (1
-\frac{V_{max}}{{E}})$ yields a family of timelike hyperbolas, see
Fig.~\ref{hyps1}. This leads to a generalization of the behavior
of orbits in region III and IV in FRW: we distinguish models that
originate from ${}^1_-\text{F}$ (resp. $(L=0)\cap (Q_0 =+1)$)
and end in ${}^1_+\text{F}$ ($(L=0)\cap (Q_0 =-1)$).

\begin{figure*}[tc]
\centering
        \subfigure[Spiral tubes that describe the
        stable manifold of a periodic orbit; on each side of the periodic orbit 25
        spiral orbits are depicted.]{
        \label{stabletube}
        \includegraphics[width=0.27\textwidth]{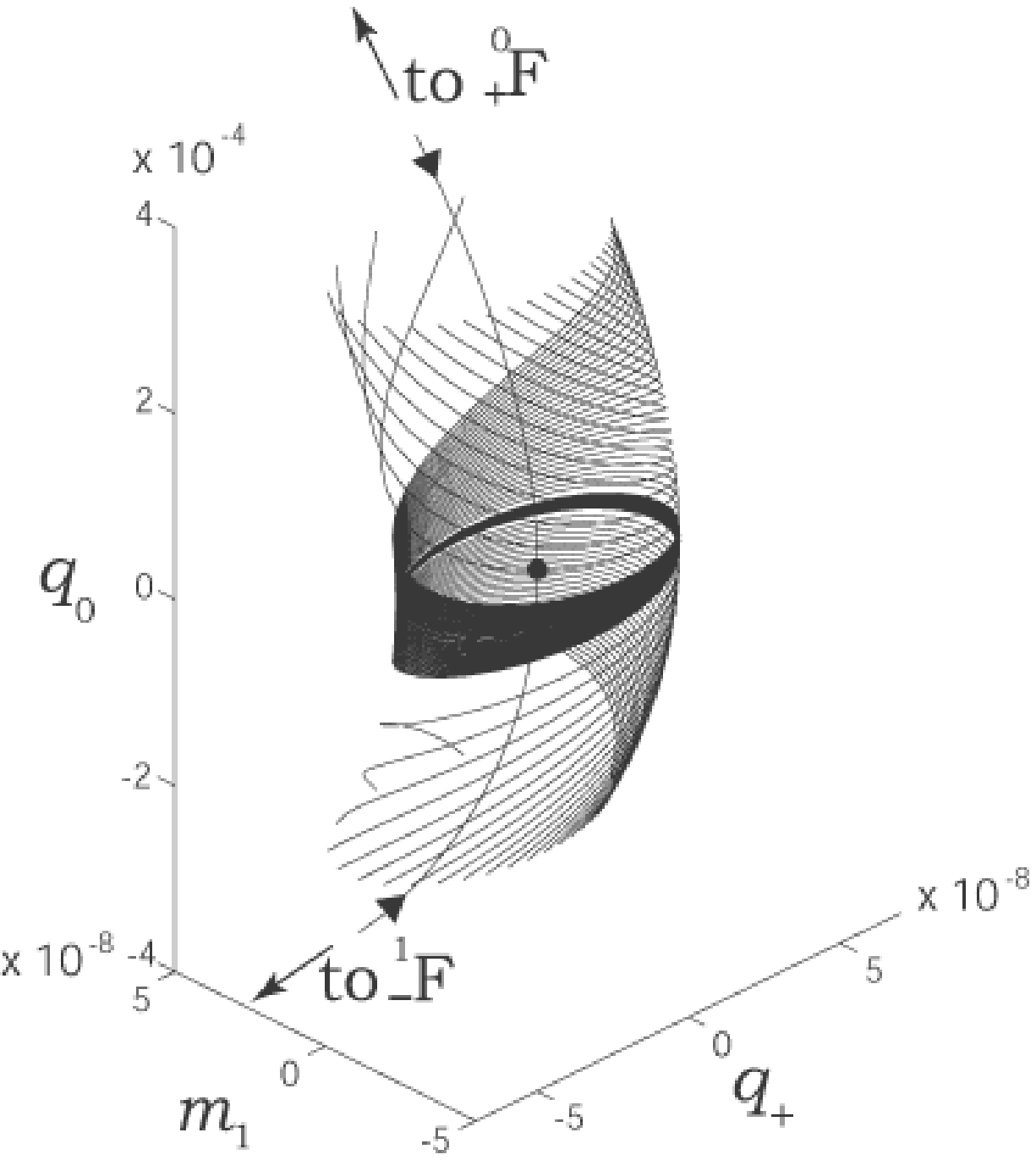}}\qquad
        \subfigure[Spiral tubes that describe the
        unstable manifold of a periodic orbit; each side of the periodic orbit consists of 25
        spiral orbits.]{
        \label{unstabletube}
        \includegraphics[height=0.30\textwidth]{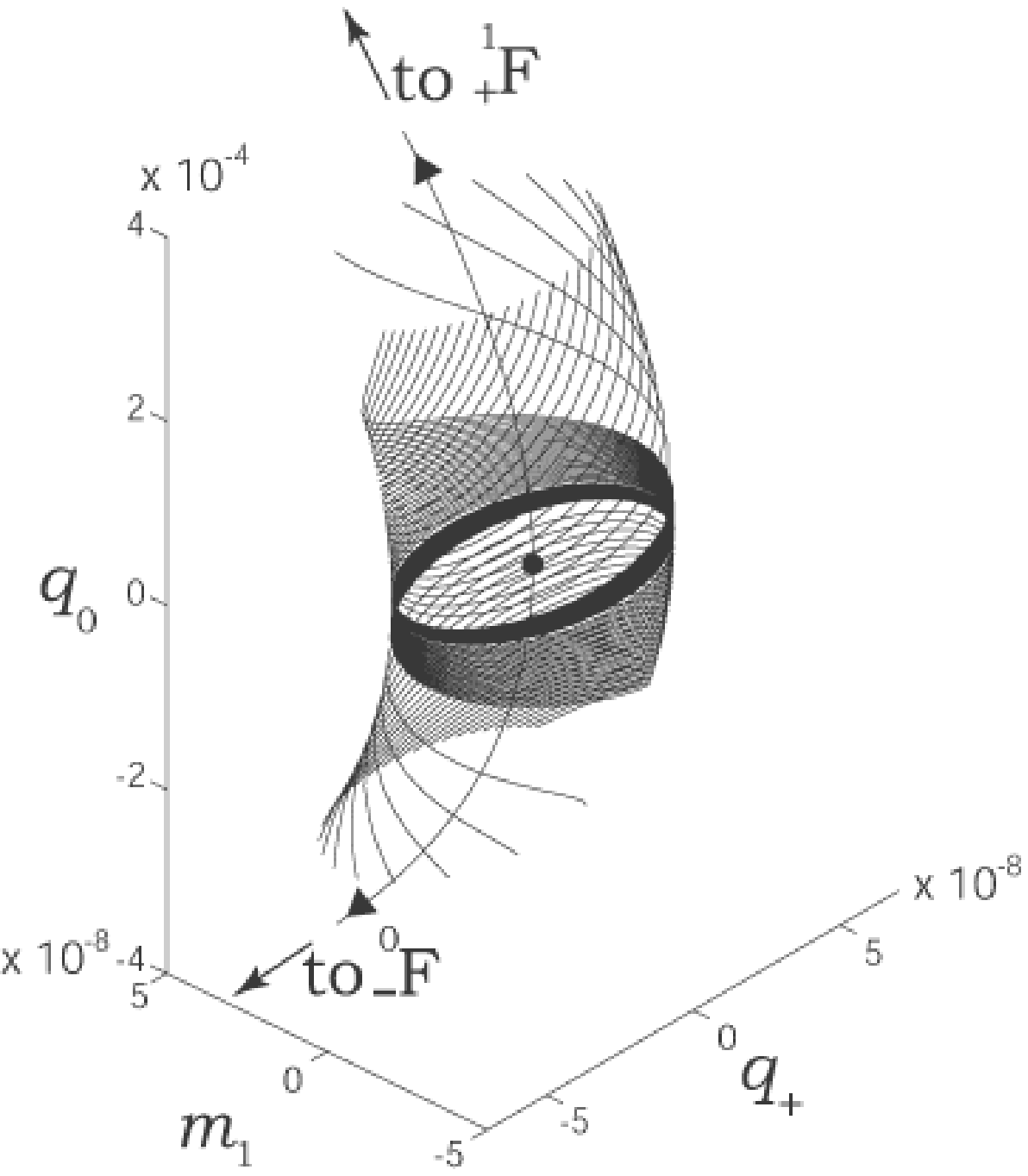}}\qquad
 \subfigure[The four spiral tubes of Figs.~(a) and (b) in the state space;
   the diameter of the tubes is too small to be resolved.
   Note that in the projection of Figs.~(a) and (b) the tubes
   would have overlapped.]{
        \label{spiralsep}
        \includegraphics[width=0.27\textwidth]{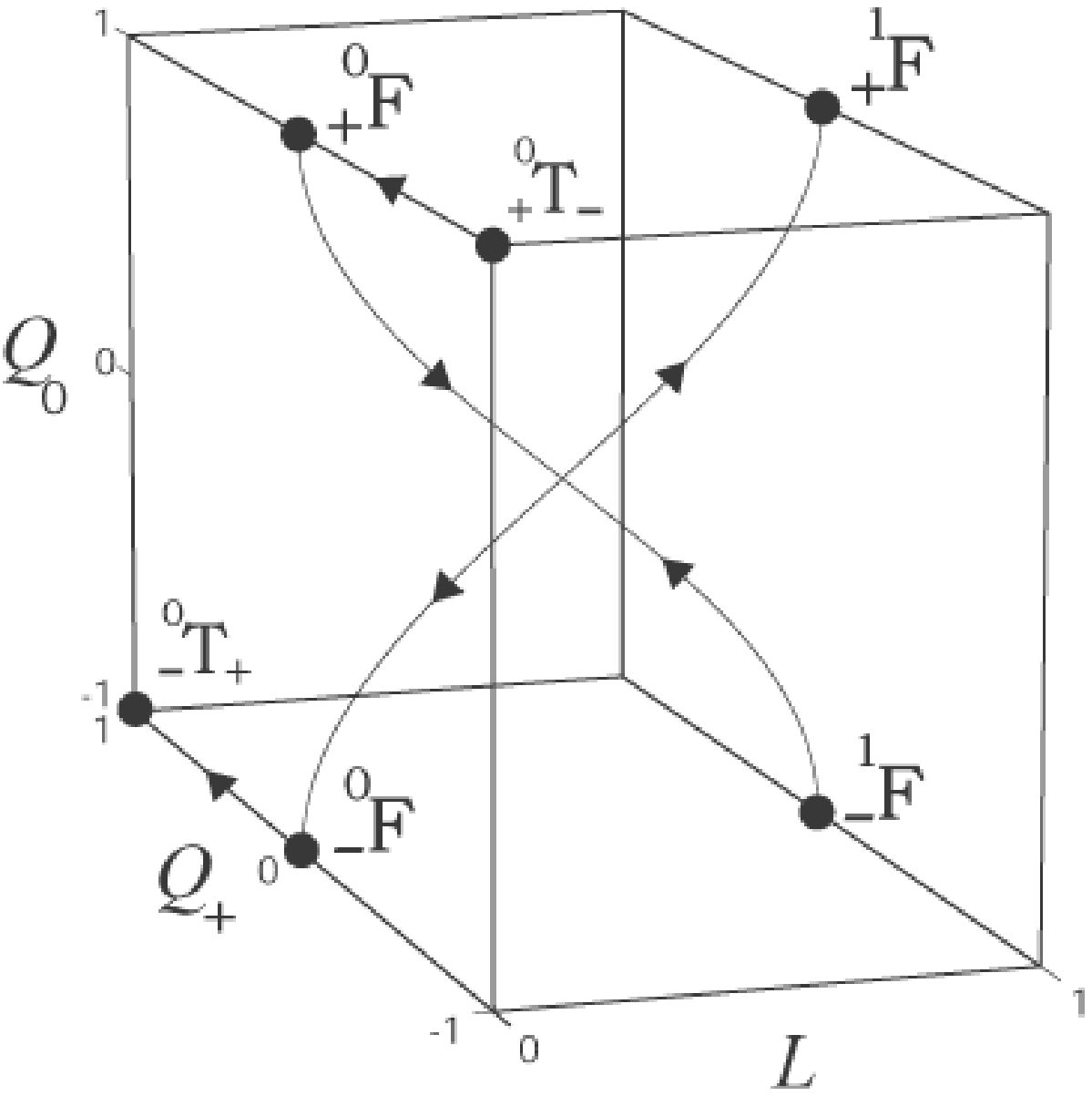}}
    \caption{Illustration of the close relationship between the dynamics close to $\text{E}$ and FRW dynamics.
      At the scale of Fig.~(c) there is no visible difference between this figure and the analogous figure depicting the FRW
      separatrices in the same projection, except that the orbits
      do not converge to ${}^0_\pm\text{F}$ in general, but to the generic limits
      ${}^0_{\mp}\text{T}_{\pm}$, cf.~App.~\ref{asystates} and Fig.~\ref{fig:Elarge}.
      In this figure the matter content is radiation and
      a cosmological constant, however, the qualitative behavior is the same for any matter source that
      admits the Einstein static solution.}
    \label{fig:EV}
\end{figure*}

\begin{figure}[tc]
\centering
  \subfigure[{\scriptsize $\|(q_+,m_1)\| < 3 (1 -\frac{V_{max}}{{E}})$}]{
        \label{hyps1}
        \includegraphics[width=0.27\textwidth]{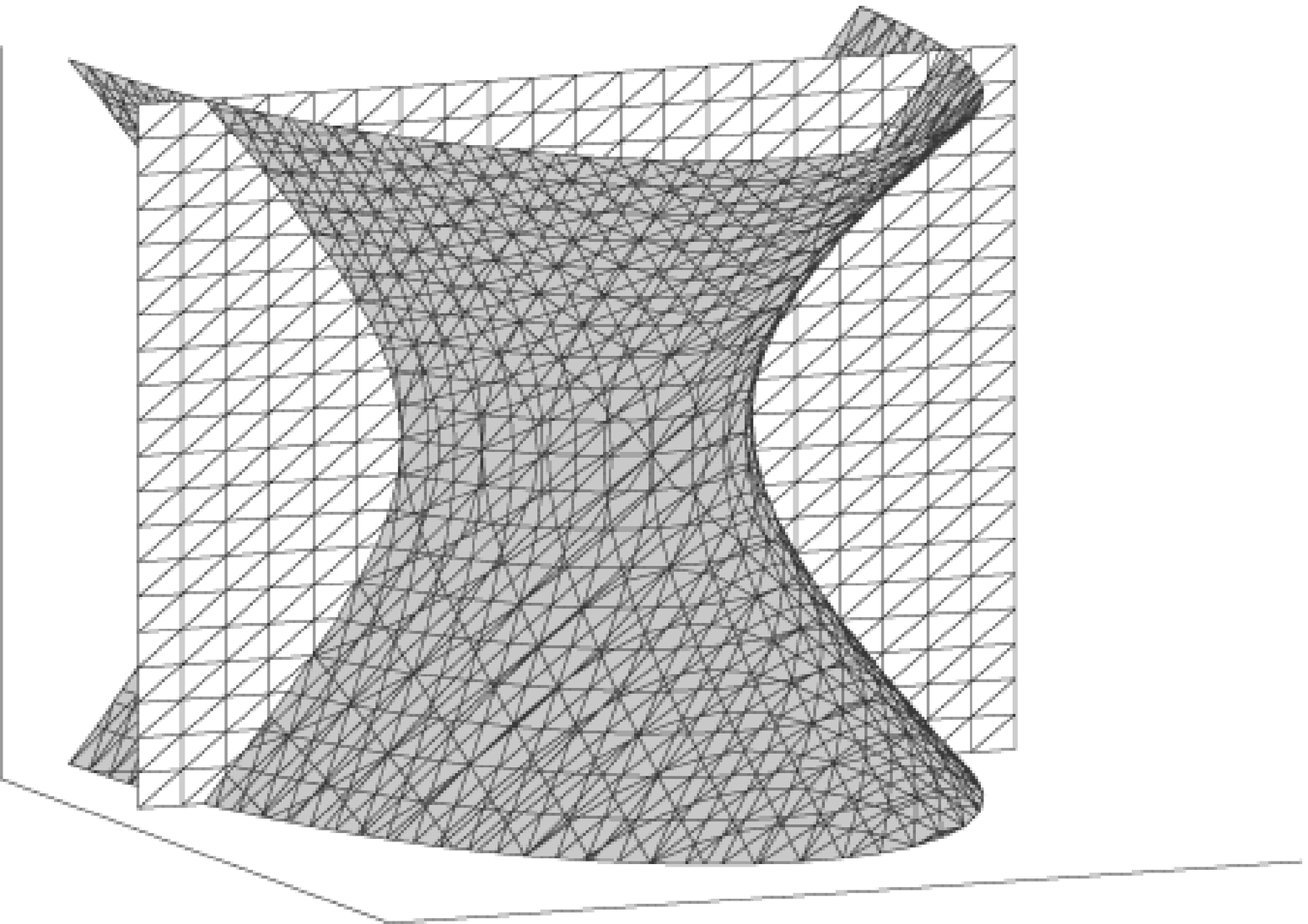}}\qquad
  \subfigure[{\scriptsize $\|(q_+,m_1)\| = 3 (1 -\frac{V_{max}}{{E}})$}]{
        \label{hyps2}
        \includegraphics[width=0.27\textwidth]{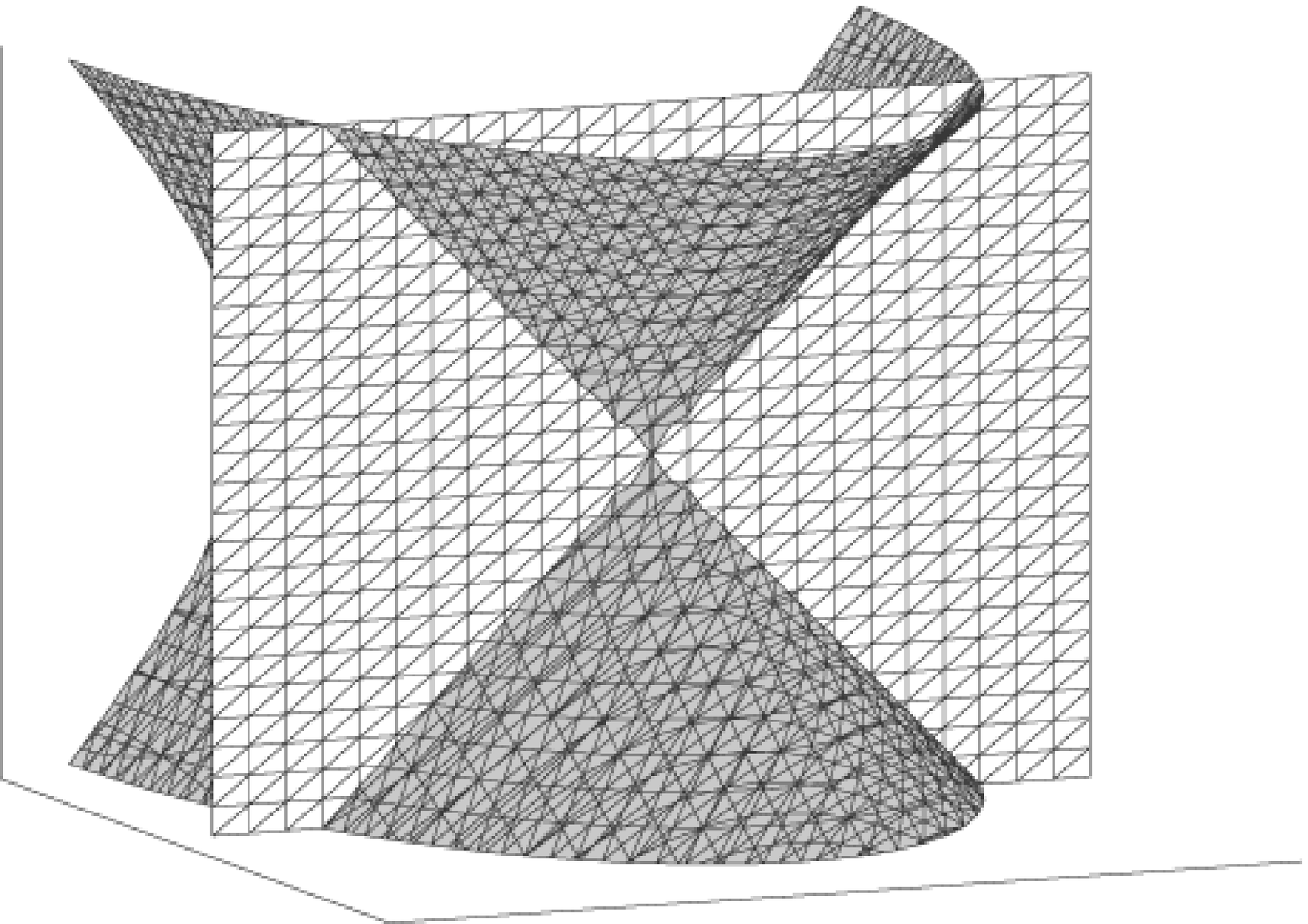}}\qquad
  \subfigure[{\scriptsize $\|(q_+,m_1)\| > 3 (1 -\frac{V_{max}}{{E}})$}]{
        \label{hyps3}
    \includegraphics[width=0.27\textwidth]{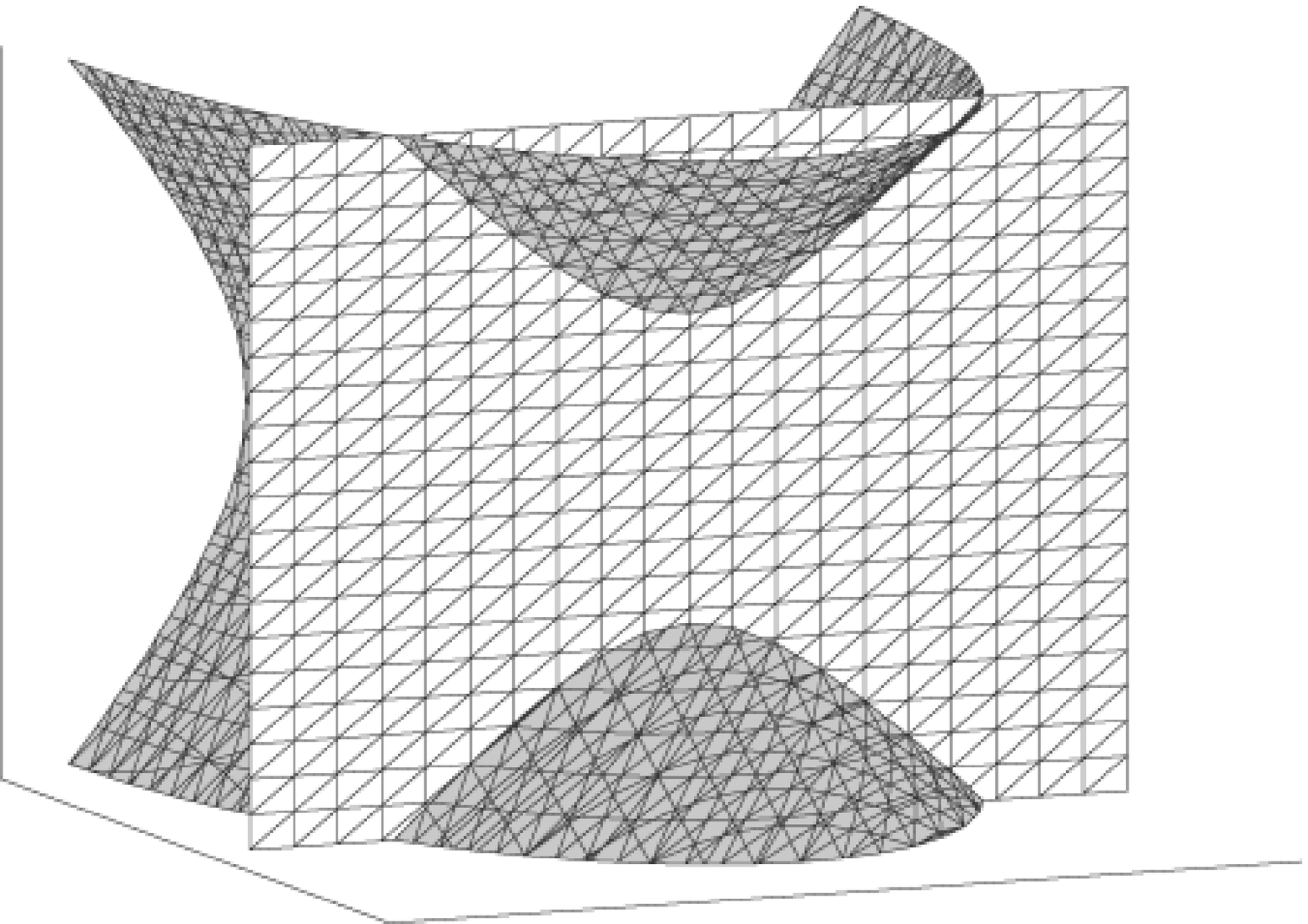}}
    \caption{Case $E<V_{max}$.
      Intersection of the timelike hyperboloids with energy $E$
      with different planes $\|(q_+,m_1)\|=const$.
      We obtain timelike hyperbolas in Case (a), a vertex and a null cone in Case (b),
      and spacelike hyperbolas in (c). The vertex of the null cone in (b) represents
      a closed orbit in the center manifold. The global asymptotic behavior of the orbits
      is governed by the null ray orbits $\tilde{\nu}^{\pm}_{l,q}$, cf.~Fig.~\ref{nullcone}.
      The axes in the figures are the same as in Fig.~(\ref{regions}).}
    \label{timehyps}
\end{figure}

Case (b): Intersection with a plane $\|(q_+,m_1)\|= 3 (1
-\frac{V_{max}}{{E}})$ yields the point $\|(q_+,m_1)\|= 3 (1
-\frac{V_{max}}{{E}})$, $q_0=0$, $l=0$, which represents a
periodic orbit, and four in/outgoing null rays (cf.~also the
discussion following Eq.~(\ref{3hyp})), see Fig.~\ref{hyps2}. Two
of these null rays represent orbits that originate from either
$(L=0)\cap (Q_0 =+1)$ or ${}^1_-\text{F}$ and converge to the
periodic orbit as $\tau\rightarrow\infty$; the other two originate
from the center manifold orbit and end at either $(L=0)\cap (Q_0
=-1)$ or ${}^1_+\text{F}$. The FRW picture is thus generalized;
periodic center manifold orbits (one for each value of $E$) take
the place of the Einstein point $\text{E}$, see Fig.~\ref{fig:EV}.
This figure also indicates how the abstract
${}^3\textbf{X}$ picture is related to the ${}^4\textbf{X}$ state space
picture, cf. Fig.~\ref{hyps1} and Fig.~\ref{fig:EV}.

Case (c): Intersection with planes $\|(q_+,m_1)\|=const> 3 (1 -
\frac{V_{max}}{{E}})$ results in a family of spacelike hyperbolas,
see Fig.~\ref{hyps3}; this case resembles the $E>V_{max}$ case and
yields similar behavior. Accordingly, there exist models that
expand forever from a singularity to infinite dispersion, despite
that their energy $E$ is lower than the ``potential barrier"
$V_{max}$. In this case $\|(q_+,m_1)\|$ acts as an additional
energy that is sufficiently large to lift the orbit over the top
of the hill from the ``side of recollapse" to the ``side of
expansion" in Fig.~\ref{figcaseII}, when it passes through the
neighborhood of $\text{E}$.

When $E$ is sufficiently close to $V_{max}$, and if the orbit
passes through a sufficiently small neighborhood of $\text{E}$,
the presented arguments about global asymptotic behavior are
mathematically rigorous. The known asymptotics of the null ray
orbits~(\ref{nullraysglobal}), and continuous dependence on
initial data, guarantees that orbits that are close to the null
rays are dragged to neighborhoods of ${}^{0,1}_{\pm}\text{F}$. The
fixed point ${}^1_{+}\text{F}$ is a sink, hence all orbits that
pass through a sufficiently small neighborhood of
${}^1_{+}\text{F}$ end in ${}^1_{+}\text{F}$. An analogous
statement holds for the source ${}^1_{-}\text{F}$. When an orbit
is dragged along $\tilde{\nu}_q^-$ towards ${}^0_-\text{F}$, it
represents a model that contracts when $L<\cl$; in
App.~\ref{globaldynamicalfeatures} we show that it must continue
to contract and reach $(L=0)\cap (Q_0=-1)$. Analogously, orbits
that head for ${}^0_+\text{F}$ along $\tilde{\nu}_l^+$ (for
$\tau\rightarrow-\infty$) expand forever from $(L=0)\cap (Q_0=1)$.

We emphasize that the dynamics of the LRS type IX models is
closely related to the FRW case and completely predictable near
the Einstein point $\text{E}$ as regards asymptotic states: in the
FRW case, the asymptotic states are determined by the energy $E$;
in the LRS type IX case, the asymptotic states are determined by
the energy $E$ and by the value of $\|(q_+,m_1)\|=const$. This
disproves the claims
of~\cite{Oliveira/etal:1997,Barguine/etal:2001} about
non-predicability of asymptotic states for models close to
$\text{E}$. Moreover, as shown in
Appendix~\ref{globaldynamicalfeatures}, the homoclinic behavior of
orbits asserted in~\cite{Oliveira/etal:1997,Barguine/etal:2001},
can be disproved not only for orbits that pass through
a neighborhood of $\text{E}$, but also for large regions
of the state space, such as $L\leq \cl$, $L > \tilde{\cl}_E$:
in these regions the asserted orbits whose $\alpha$- and
$\omega$-limit is the same periodic orbit cannot exist.
Furthermore, numerical results reveal that the qualitative
features described above for a sufficiently small neighborhood of $\text{E}$
carry over to a large neighborhood of $\text{E}$ and a large range of
energies, as can be seen in Fig.~\ref{spiral} and
Fig.~\ref{fig:Elarge}. This indicates that the assertions about
``homoclinic phenomena and chaos'' are wrong for the entire state space.

\begin{figure}[tc]
\centering
  \subfigure[Periodic orbits act as ``saddle" orbits in the neighborhood of E.]{
        \label{EA}
        \includegraphics[width=0.26\textwidth]{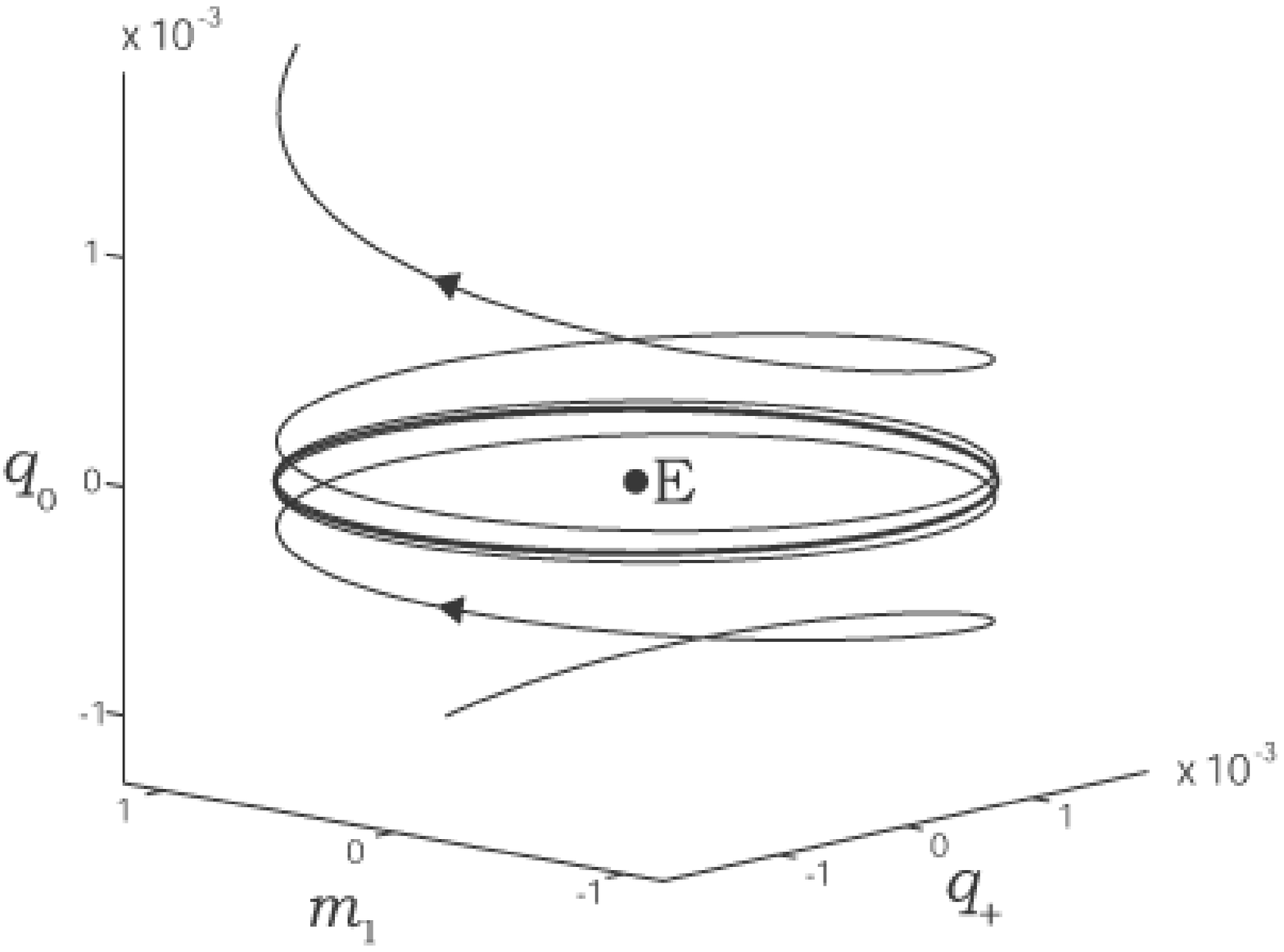}}\quad
  \subfigure[The center manifold dominates even far away from E.]{
        \label{EB}
        \includegraphics[width=0.26\textwidth]{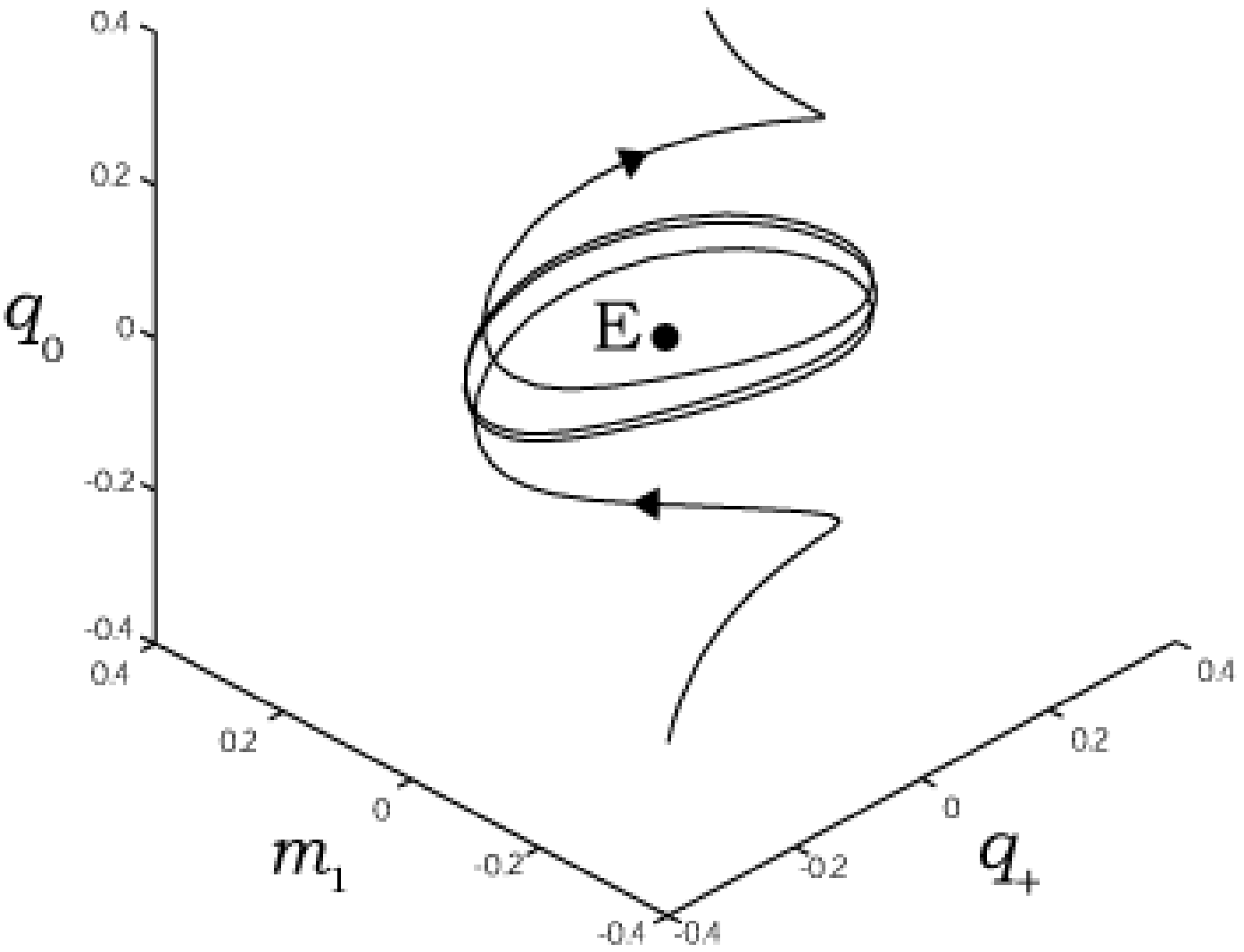}}\quad
  \subfigure[The same orbit as in (b), now depicted in the state space through
        a $(Q_0,Q_+,M_1)$-projection.]{
        \label{EC}
    \includegraphics[width=0.32\textwidth]{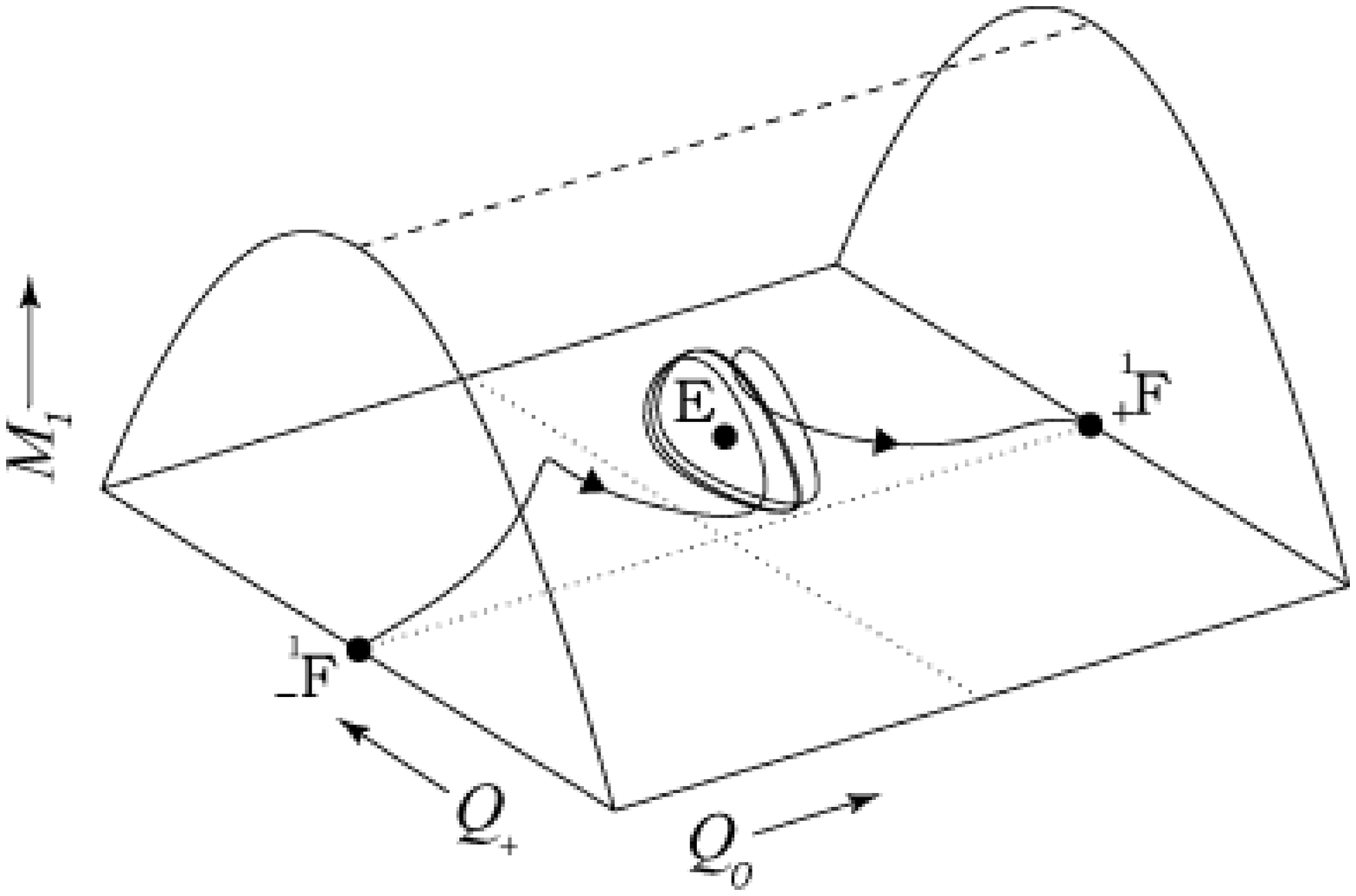}}
    \caption{Example of the effect of the center manifold on dynamics. The source is
    orthogonal dust and a positive cosmological constant, however, the pictures
    depict behavior that is typical for all sources that admit a unique Einstein point $\text{E}$.}
    \label{spiral}
\end{figure}

\begin{figure*}[tc]
\centering
        \subfigure[Spiral tubes that emanate from a periodic orbit far from the Einstein point.
        On each side of the periodic orbit, 25 orbits with the same energy that describe the
        stable manifold of the periodic orbit, are depicted.]{
        \label{stabletube_large}
        \includegraphics[width=0.33\textwidth]{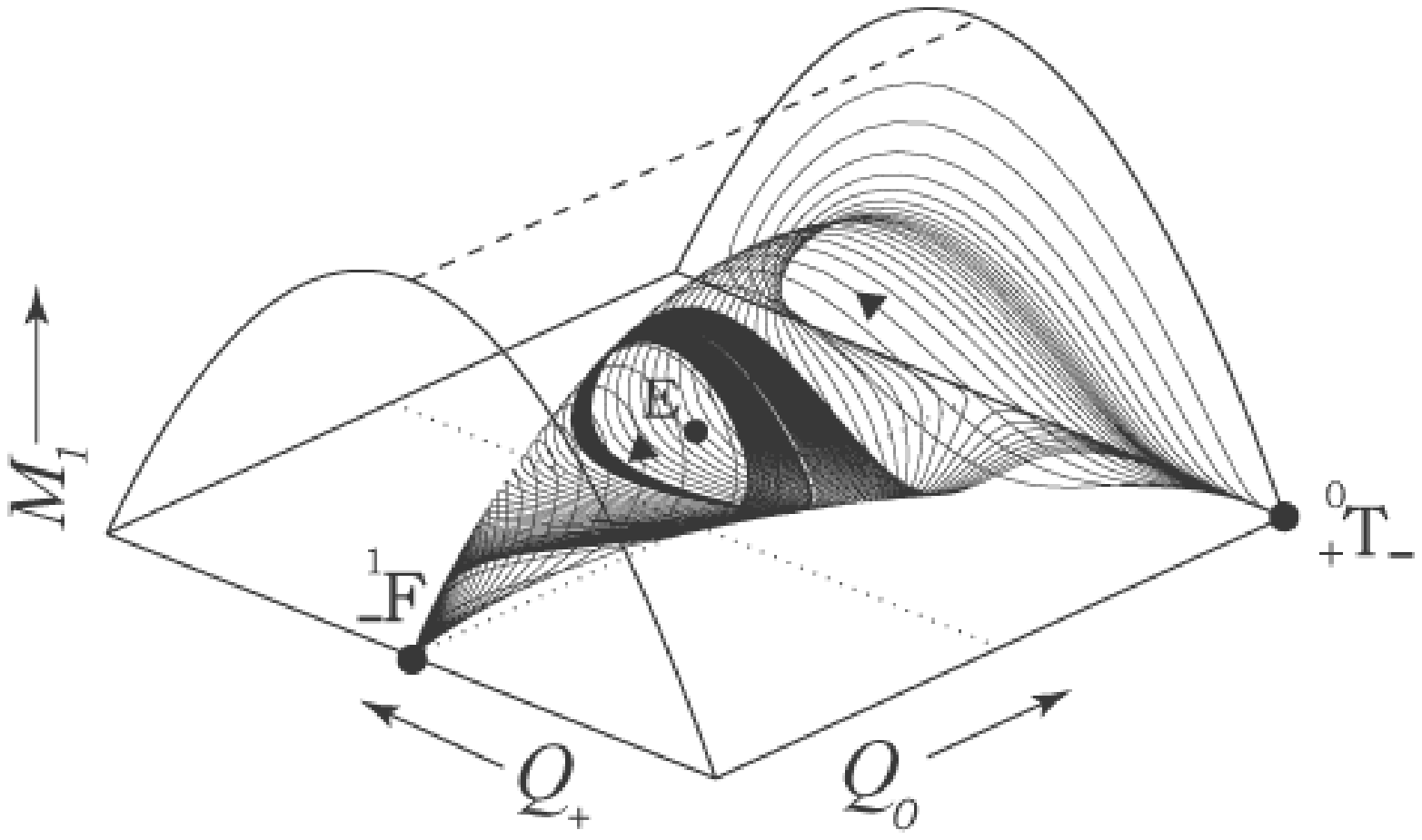}}\qquad
        \subfigure[Spiral tubes that emanate from a periodic orbit far from the Einstein point.
        On each side of the periodic orbit, 25 orbits with the same energy that describe the
        unstable manifold of the periodic orbit, are depicted.]{
        \label{unstabletube_large}
        \includegraphics[height=0.20\textwidth]{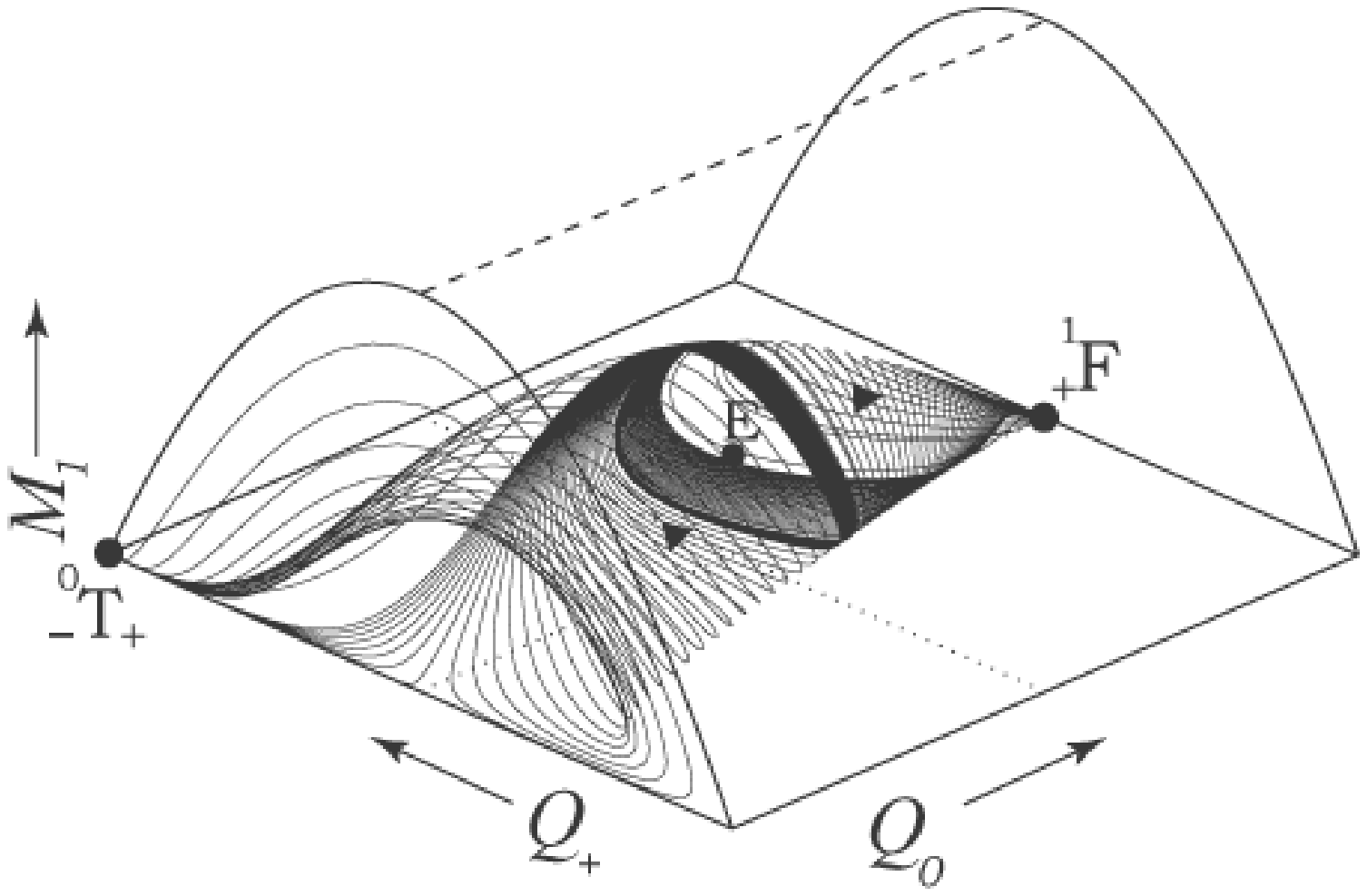}}\qquad
 \subfigure[The four spiral tubes of Figs.~(a) and (b) in the state space;
   they emanate from the stable Friedmann/Taub points
   on the expanding and contracting Bianchi type II
   boundary subsets.]{
        \label{spiralcross}
        \includegraphics[width=0.24\textwidth]{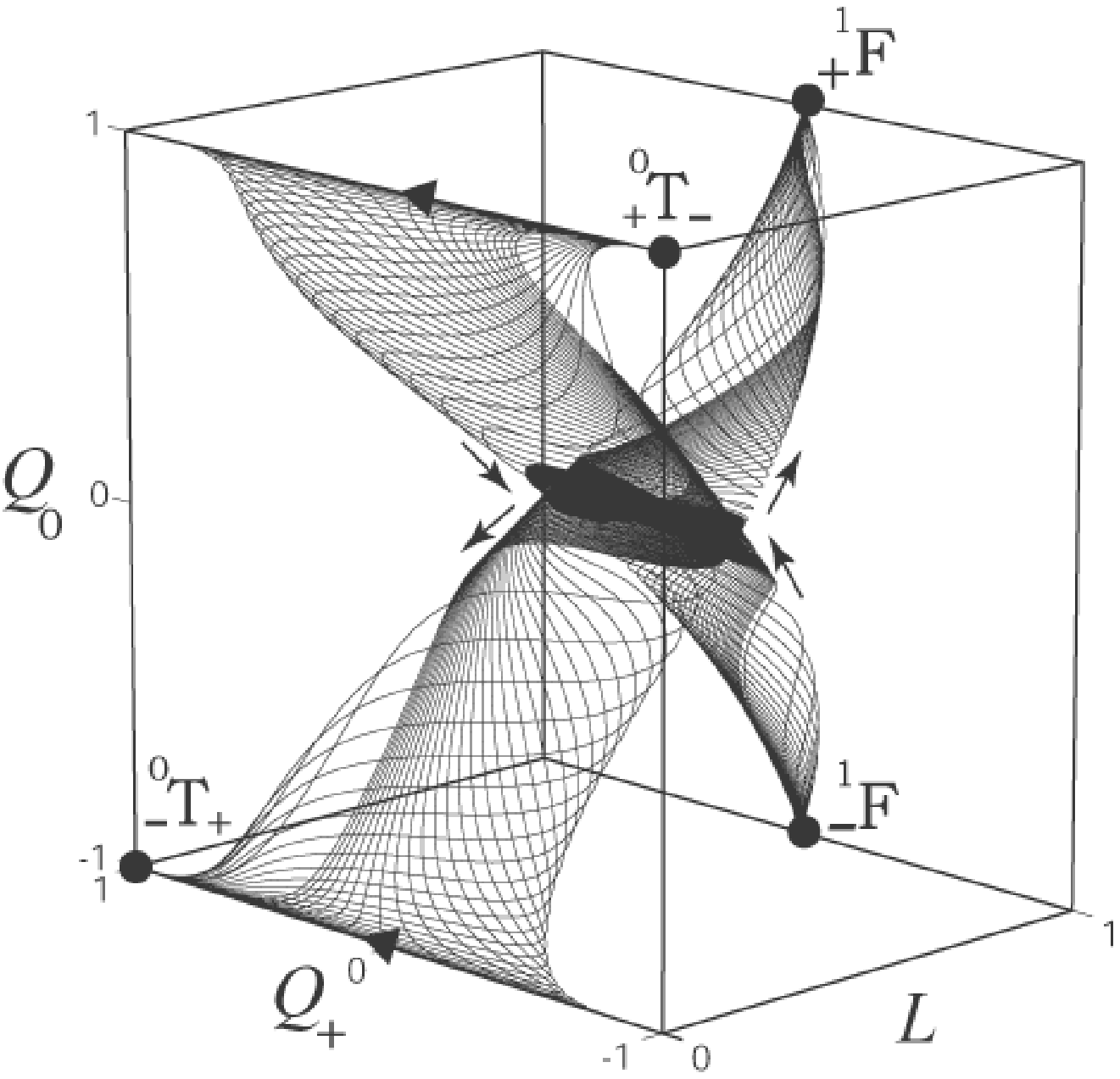}}
    \caption{Illustration of the dynamics far from
      the Einstein point: the qualitative structure of the flow is still the one of
      a small neighborhood of $\text{E}$. Clearly there is no
      ``homoclinic chaos." In this figure the matter content is radiation and a
      cosmological constant, however, the qualitative behavior is the same for
      any matter source that admits the Einstein static solution.}
    \label{fig:Elarge}
\end{figure*}

Finally note that the discussion of the dynamics in the
neighborhood of the Einstein point $\text{E}$ is independent of
the assumption of asymptotic linearity of the equation of state
and that the global asymptotic behavior of orbits remains valid in
a wide sense. The boundaries are no longer part of the systems,
however, it is still correct to state that the $\alpha$- and
$\omega$-limits of orbits are located on $(L=0) \cap (Q_0 =\pm 1)$
and $(L=1)\cap (Q_0 =\pm 1)$.

\section{Kantowski-Sachs cosmologies}
\label{KSmodels}

\begin{figure}[ht]
\centering
\includegraphics[height=0.28\textwidth]{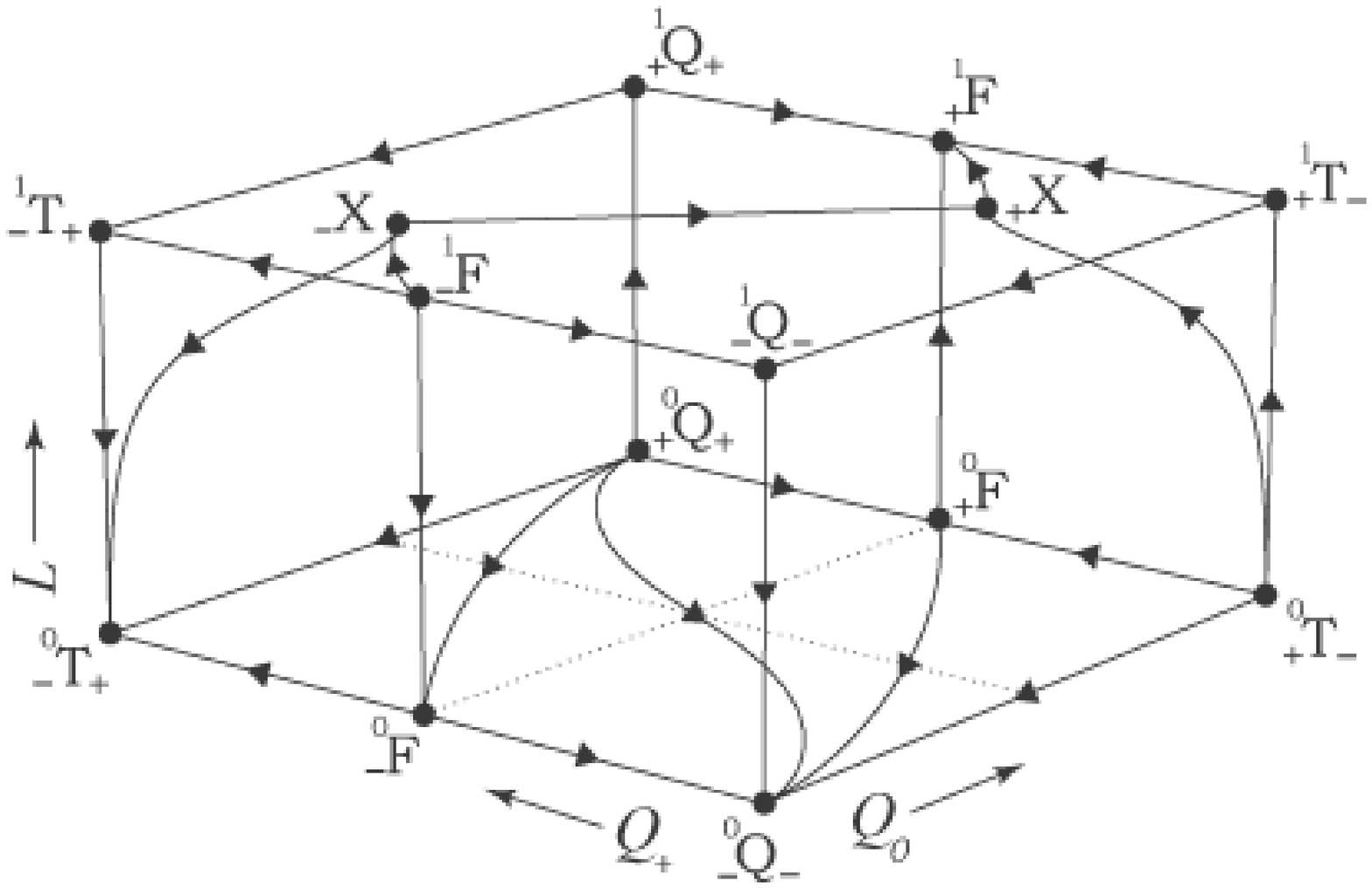}
\caption {The Kantowski-Sachs submanifold. Case (ii).}
\label{fig:L0tent}
\end{figure}

The coupled system of equations for the Kantowski-Sachs models is
obtained from the general system~(\ref{dynsys}) by setting
$M_1=0$; we thus have the three-dimensional state space
$\{(Q_0,Q_+,L)\}$ depicted in Fig.~\ref{fig:L0tent}. The coupled
system admits a number of equilibrium points whose associated
eigenvalues are given in Table~\ref{tab4DfixpointsKS}.

\begin{table}[ht]
  \begin{center}
    \begin{tabular}{c|ccc}

      F.P & $\lambda_1$ & $\lambda_2$ & $\lambda_3$ \\  \hline
    & & &    \\[-0.3cm]
      $^A_\pm$F           & $\pm(1+3w_A)$ & $\mp\frac{3}{2}(1-w_A)$ & $ \pm S(A)f_A $ \\
      ${^A_\pm}$Q$_{\pm}$ & $\pm$2 & $\pm 3(1-w_A)$ & $\pm S(A)f_A$ \\
      ${^A_\pm}$T$_{\mp}$ & $\pm 6$ & $\pm 3(1-w_A)$ & $\pm S(A)f_A$ \\
      $_{\pm}$X           & $\mp\frac{2f_1}{1-3w_1}$ & $\mp\frac{3}{2}\frac{1-w_1}{1-3w_1}\left(1+\sqrt{d}\right)$
                          & $\mp\frac{3}{2}\frac{1-w_1}{1-3w_1}\left(1-\sqrt{d}\right)$ \\

    \end{tabular}
  \end{center}
    \caption{Fixed points and eigenvalues for the Kantowski-Sachs submanifold.
             Here $A=0,1$, $S(0)=1$, $S(1)=-1$,
             $d=(24w_1^2+7w_1+1)/(1-w_1)$.}
    \label{tab4DfixpointsKS}
\end{table}

It follows that in Case (i) all fixed points are saddles except
for $_+^0\text{T}_-$, which is a source, and $_-^0\text{T}_+$,
which is a sink. In Case (ii) all fixed points are saddles except
for $_+^0\text{T}_-$ and $_-^1\text{F}$, which are sources, and
$_-^0\text{T}_+$ and $_+^1\text{F}$, which are sinks.

In Case (i) all models expand from a singularity, reach a point of
maximum expansion, and then recollapse to a singularity, as
follows from an analysis similar to that of the LRS type IX case.
The situation in Case (ii) is more complicated; local analysis
limits the possibilities somewhat but numerics is required to give
a detailed picture. Fig.~\ref{fig:L0tent} illustrates various
possibilities, e.g., there are solutions that expand from
singularities to infinitely dispersed isotropic states and
solutions that contract from infinitely dispersed isotropic states
to singularities.

\section{Conclusion and outlook}
\label{discandout}

In this paper we have regularized Einstein's field equations and
obtained a dynamical system on a compact state space for the LRS
Bianchi type IX and Kantowski-Sachs perfect fluid models with a
barotropic equation of state. This allowed us to systematically
study the effects of matter properties on cosmological evolution,
and we obtained a global picture of the dynamical possibilities.
In type IX, the cornerstones of our analysis were: center manifold
theory in connection with Einstein's static solution --when
existent; an ``energy integral" that severely restricted the
dynamics; a close connection between the FRW and LRS type IX
cases. This led to a comprehensive description of the dynamics in
the neighborhood of Einstein's static model: given an initial
value in the state space in the neighborhood of Einstein's model,
it is possible to predict whether the solution will expand to an
infinite dilute final state or eventually contract to a
singularity. This disproves previous assertions that the dynamics
near the Einstein model exhibits non-predictability and chaos. The
present formalism is generalizable to cover more general models
such as general Bianchi type IX models for which it is also
possible to derive a conserved energy which, together with center
manifold theory, again severely restrict the dynamics; indeed, we
expect that the conclusions drawn in this paper will have direct
bearing on this problem as well. In addition to this we have
obtained several global results; again we expect our methods and
results to be generalizable to the general Bianchi type IX case.

The present approach is also of relevance for the forever
expanding Bianchi type I--VIII perfect fluid models. As in the
present case one can introduce a variable $L$ and obtain an
autonomous formulation, however, in this case one can also use
$\ln (\ell/\ell_0)$ as a time variable (the natural time variable
when $H$ is used as the normalization factor, as discussed in
\cite{book:waiell97}), which then leads to that $w$ becomes a time
dependent function and thus that one obtains a non-autonomous
problem. However, in the case of an asymptotic linear equation of
state one obtains an asymptotically autonomous problem and for
certain problems one can then apply the Strauss-Yorke theorem to
determine the asymptotic dynamics \cite{stryor1967},
\cite{horetal2003}. Both approaches are expected to have
advantages and disadvantages that depend on the problem at hand.
Although one can draw the conclusion that Bianchi type I--VIII
perfect fluid models with $\rho>0$ are forever expanding, the
details of how this actually happens is a quite intricate problem
due to asymptotic self-similarity breaking, which occur for the
more general classes of models \cite{waietal1999},
\cite{horetal2003}; we expect asymptotic self-similarity breaking
to hold for the same class of models also for asymptotically
linear equations of state, but the details will depend on the
equation of state.

We thus expect that some of the ideas in this paper are
generalizable to other SH models, however, we also believe that
some ideas are generalizable to models with fewer symmetries, even
to models with no symmetries at all, so-called $G_0$ models,
particularly when it comes to asymptotic dynamics associated with
singularity formation. Again one can treat a barotropic equation
of state by introducing a function similar to $L$ that parametrize
the equation of state suitably. In special cases of matter
dominated singularities additional assumptions like asymptotic
linearity will have to be imposed on the equation of state,
however, generically the singularity is expected to be ``vacuum
dominated" \cite{uggetal2004}; in this case we expect that one can
treat equations of state with quite general asymptotic behavior.

\subsection*{Acknowledgements}

We would like to thank Alan Rendall for comments on an earlier
draft. C.U. was in part supported by the Swedish Research Council.


\begin{appendix}

\section{Dynamical systems nomenclature and definitions}
\label{nomenclature}

Consider a dynamical system $\dot{y}= f(y)$ and the associated
flow $\Phi_t$. The $\omega$-limit $\omega(x)$ ($\alpha$-limit
$\alpha(x)$) of a point $x$ is defined as the set of all
accumulation points of the future orbit $\{\Phi_t(x)\,|\, t>0\}$
(past orbit $\{\Phi_t(x)\,|\, t<0\}$) of $x$. The $\omega$-limit
of a set $S$ is $\omega(S) = \bigcup_{x\in S}\omega(x)$. An orbit
$\{\Phi_t(x)\,|\,t\in\mathbb{R}\}$ is called homoclinic, if there
exists a (fixed) point $z$ such that $\alpha(x) =\omega(x) =
\{z\}$; an orbit is called heteroclinic if it originates from a
fixed point $z_1$ and ends in a different fixed point $z_2$. A set
$S$ is called future (past) invariant, if $\Phi_t(S) \subseteq
S\:\forall\, t>0$ ($\forall\, t<0$).

The monotonicity principle states that if there is a function $Z$
on an invariant set $S$ that is strictly decreasing along orbits,
then $\omega(S) =\{ s\in \bar{S}\backslash S\,|\,
\lim_{x\rightarrow s} Z(x) \neq \sup_S Z\}$, and analogously for
the $\alpha$-limit.

Two dynamical systems are called (locally) equivalent, if there
exists a homeomorphism (of coordinates) $\Psi$ such that the flows
are topologically equivalent $\Phi_t^1 = \Psi^{-1} \circ \Phi_t^2
\circ \Psi$. Under certain conditions the equivalence is a
diffeomorphism.

\section{Asymptotic states and global dynamics}
\label{globaldynamicalfeatures}

In this section we discuss aspects of the global dynamics of case
(ii) of the LRS type IX models: first we characterize the
asymptotic states of models that converge to a singularity or to
an infinitely diluted state, then we describe the global
qualitative behavior of models with $E\geq V_{max}$, and thirdly
we treat some aspects of the global dynamics of models with
$E<V_{max}$; the considerations complement the investigation of
Sec.~\ref{w-1--1/3} which focussed on the dynamics near the
Einstein point $\text{E}$.

\subsection{Asymptotic states}
\label{asystates}

To investigate the asymptotic behavior of orbits that converge to
a singularity as $\tau\rightarrow\infty$ we make use of the
conserved energy~(\ref{genint}). In analogy
with~(\ref{limitof1-Q02}) we find $Q_0 \rightarrow -1$ as
$\tau\rightarrow \infty$, i.e., the attractor is the set
$(L=0)\cap (Q_0 =-1)$. In the FRW picture~\ref{FRWsub} the
attracting surface $(L=0)\cap (Q_0 =-1)$ reduces to the fixed
point ${}^0_-\text{F}$; in the general picture $(L=0)\cap (Q_0
=-1)$ contains four fixed points: ${}^0_-\text{F}$,
${}^0_{-}\text{Q}_{-}$, ${}^0_{-}\text{T}_{+}$, and
$^0_-\text{CS}$. The generic attractor is ${}^0_{-}\text{T}_{+}$,
a three-parameter family of orbits ends there; ${}^0_-\text{F}$
attracts a two-parameter set of orbits, $^0_-\text{CS}$ a
one-parameter set, cf.~Table~\ref{tab4Dfixpoints}. Since no
periodic orbits or heteroclinic cycles exist on $(L=0)\cap (Q_0
=-1)$ (see \cite{book:waiell97}), the fixed points are the only
possible $\omega$-limits. Analogously, the surface $(L=0)\cap (Q_0
=1)$ contains the $\alpha$-limit of orbits that expands from a
singularity; the treatment of the fixed points is analogous.

When the equation of state is not asymptotically linear, i.e.,
when $w\not\rightarrow w_1$, but $\inf w > -1/3$ as $L\rightarrow
0$, then $L=0$ is not included in the state space and the local
analysis of the fixed points breaks down; however, since
${}^0_{-}\text{T}_{+}$ is a sink for all $-1/3 < w_1 < 1$, the
point $(-1,+1,0,0)$ remains a generic attractor of a
three-parameter family of orbits also in the general case.

The asymptotic behavior of orbits that expand to a state of
infinite dispersion is qualitatively simpler. As
$\tau\rightarrow\infty$ the orbit approaches the attracting set
$(L=1)\cap (Q_0=1)$; to see that we use the same arguments as
above. However, on $(L=1)\cap (Q_0=1)$, among the fixed points
${}^1_+\text{F}$, ${}^1_{+}\text{Q}_{+}$, ${}^1_{+}\text{T}_{-}$,
it is only the point ${}^1_+\text{F}$ that attracts orbits from
the interior of the state space, see Table~\ref{tab4Dfixpoints}.
Analogously, ${}^1_-\text{F}$ is the only source on the set
$(L=1)\cap (Q_0=-1)$.
These statements remain valid also when the equation of
state is not asymptotically linear.

\subsection{$\mathbf{E \geq V_{max}}$}
\label{EgeqVmax}

We prove that models that satisfy $E > V_{max}$, i.e., models
in regions I and II, are forever expanding or forever contracting
between a singularity and a state of infinite dilution.
Consider a solution $(Q_0,Q_+,M_1,L)(\tau)$ with $E>V_{max}$ and
consider~(\ref{genint}) divided by $V_{max}$,
\begin{equation}\label{EbiggerVmax}
\frac{E}{V_{max}} = \frac{3}{4^{4/3}} \Omega_D^{-1}
(1-Q_0^2)^{4/3} M_1^{-1/3} \,\frac{V(L)}{V_{max}}\:.
\end{equation}
Since $V(L)/V_{max} \geq 1$ and by assumption $E/V_{max} < 1$ it
follows that
\begin{equation}
\frac{3}{4^{4/3}} (1-Q_0^2)^{4/3} < \Omega_D M_1^{1/3} \leq
\frac{3}{4^{4/3}}\:,
\end{equation}
where $\Omega_D M_1^{1/3}$ has been bounded by its maximal value.
We therefore conclude that $Q_0^2 > const^2 \:\,\forall\, \tau$, hence
either $Q_0(\tau)>const$ or $Q_0(\tau)<-const \:\,\forall\, \tau$. From the
system~(\ref{dynsys}) it follows that $L^\prime > const\, f(L) L (1-L)$ or
$L^\prime < -const\, f(L) L (1-L)$ and the claim is established.

Let us juxtapose the above argument with the discussion in
Sec.~\ref{w-1--1/3}. Recall that $E>V_{max}$ corresponds to a pair
of mass hyperboloids in the neighborhood of the Einstein point
$\text{E}$. Thus, locally, it follows that $Q_0$ ($= q_0$) is
either always positive or always negative when $E>V_{max}$, hence
$L^\prime < 0$ or $L^\prime >0$; the above argument proves that
the mass hyperboloids extend to hypersurfaces that satisfy $Q_0>0$ or
$Q_0<0$ globally in the state space.

Models that satisfy $E = V_{max}$ are forever expanding or forever
contracting between a singularity and a state of infinite
dilution; however, five models are exceptions: the Einstein model
and the four FRW models that correspond to the separatrix orbits
-- the latter are forever expanding or contracting but converge to
$\text{E}$. To prove the claim consider solutions
$(Q_0,Q_+,M_1,L)(\tau)$ with $E=V_{max}$; from~(\ref{EbiggerVmax})
we obtain
\begin{equation}\label{EequalVmax}
\frac{3}{4^{4/3}} (1-Q_0^2)^{4/3} = \Omega_D M_1^{1/3} \frac{V_{max}}{V(L)} < \frac{3}{4^{4/3}}
\end{equation}
for all $L\neq \cl$. ($\Omega_D M_1^{1/3}$ attains
its maximal value $3/4^{4/3}$ at $(Q_+,M_1) = (0,1/4)$.)
Eq.~(\ref{EequalVmax}) implies $Q_0(\tau) \gtrless 0$ and hence $L^\prime(\tau) \gtrless 0$
when $L(\tau) \neq \cl$.
Assume that there exists $\tau_0$ such that $L^\prime(\tau_0) = 0$.
It follows that $L(\tau_0) = \cl$ and $Q_0(\tau_0) = 0$;
via the equation in~(\ref{EequalVmax}) we obtain $(Q_+(\tau_0), M_1(\tau_0)) = (0,1/4)$, i.e.,
$(Q_0,Q_+,M_1,L)(\tau_0) = \text{E}$, hence the solution coincides with the
Einstein model.
Therefore, except for the Einstein universe, all models with $E=V_{max}$
are forever expanding or contracting (so that $L(\tau)$ is strictly monotonic).
Assume that a model satisfies $L(\tau)\rightarrow L_0\neq 0,1$ for $\tau\rightarrow \pm\infty$;
then $L^\prime(\tau)\rightarrow 0$ (at least for a subsequence $(\tau_n)_{n\in\mathbb{N}}$)
and $Q_0(\tau)\rightarrow 0$ as $\tau\rightarrow \pm\infty$;
from~(\ref{EequalVmax}) we derive that $L_0 = \cl$
and $Q_+(\tau) \rightarrow 0$, $M_1(\tau)\rightarrow 1/4$.
Therefore, the model has $\text{E}$ as an $\alpha$- or $\omega$-limit,
so that it must correspond to one of the four FRW separatrix orbits, cf.~Sec.~\ref{w-1--1/3}.
We conclude that apart from the five special cases
all models with $E=V_{max}$ satisfy $L\rightarrow 0,1$ for $\tau\rightarrow \pm\infty$,
and the claim is established.

\subsection{$\mathbf{E < V_{max}}$}

Models with $E<V_{max}$ exhibit more intricate behavior as already
seen in the discussion of Sec.~\ref{w-1--1/3};
however, chaotic dynamics is excluded from the outset
in large parts of the state space, as we will see in the following.

Models that contract for some $L\leq \cl$ continue to contract until
they reach a singularity. Similarly, models that expand for some
$L\leq \cl$ must have expanded from a singularity. The
proof is analogous to that in Sec.~\ref{wbigger-1/3}: when
$L<\cl$, the surface $Q_0=0$ is a semipermeable membrane,
$Q_0^\prime <0$ on $Q_0 =0$. Thus, $(Q_0 < 0)\cap (L < \cl)$ is a
future invariant set with $L^\prime < 0$, while
$(Q_0 > 0)\cap (L < \cl)$
is a past invariant set with $L^\prime >0$; taking the
structure of the flow on $Q_0=0$ into account, the
claim follows.
When $L=\cl$, $Q_0^\prime \leq 0$ still holds; in fact one can
show that $Q_0^\prime < 0$ when $Q_+\neq 0$;
when $Q_+ =0$ we have $Q_0^\prime = Q_0^{\prime\prime} =0$, but
$Q_0^{\prime\prime\prime} = -4 (1-4 M_1)^2$, which is negative everywhere
except for at the Einstein point. Hence
$Q_0 = 0$ is semipermeable for $L\leq \cl$ and the claim is established.

From the above it follows that all models that pass through $Q_0=0$
when $L<\cl$ are expanding--contracting, i.e., these models
expand from a singularity, reach a point of maximum expansion, and
re-contract to a singularity.

\begin{figure}[hp]
\psfrag{a}[cc][cc]{{\tiny $-1$}} \psfrag{b}[cc][cc]{{\tiny $+1$}}
\psfrag{E}[cc][cc]{{\scriptsize $\ce = E$}}
\psfrag{F}[cc][cc]{{\scriptsize $F = 0$}}
\psfrag{Q}[cc][cc]{{\scriptsize $Q_+$}}
\psfrag{M}[cc][cc]{{\scriptsize $M_1$}}
 \centering
\includegraphics[width=0.35\textwidth]{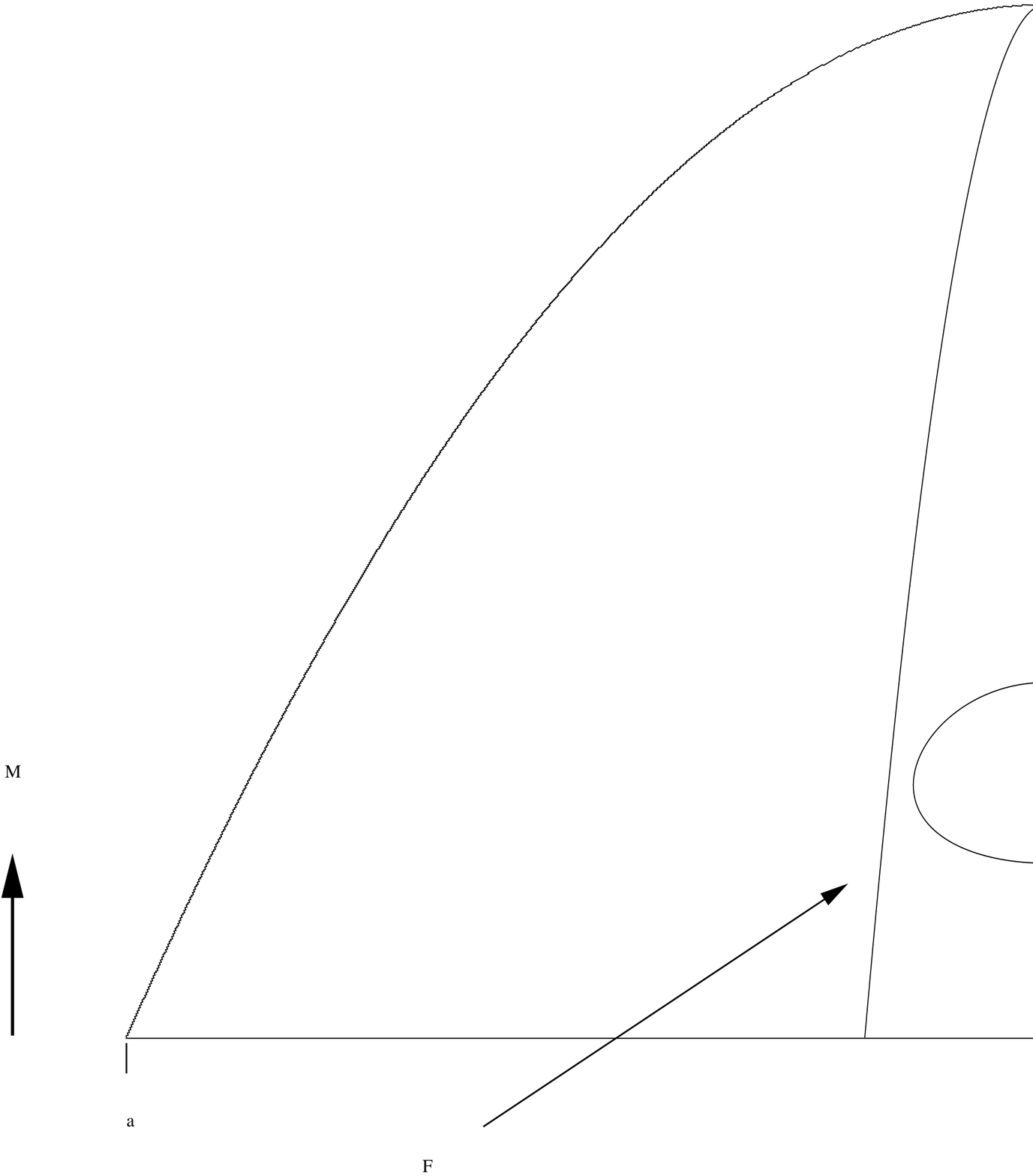}
\caption{In $Q_0 =0$, for $L =const > \tilde{\cl}_E$,
the curve $\ce=E$ is entirely contained in the region
$F < 0$ where $Q_0^\prime > 0$.}
\label{semifig}
\end{figure}

A model with energy $E < V_{max}$ that expands for some $L> \cl_E$
continues to expand forever to a infinitely diluted state. Similarly,
a model that contracts for some $L> \cl_E$ must have contracted
throughout its past history. Here, $\cl_E$ is uniquely defined
by $V(\cl_E) = E$, $\cl_E > \cl$;
note that $\cl_E \rightarrow \cl$ when $E\rightarrow V_{\max}$.
The proof of the claim is a variation of the arguments used in~\ref{EgeqVmax}:
consider the hyperplane $Q_0 =0$; in $Q_0 =0$,
the surface $\ce = E$ is given by the set of all
$(Q_+, M_1, L)$ such that $E/V(L) = (3/4^{4/3})\, \Omega_D^{-1} M_1^{-1/3}$.
For each $L =const$ with $\cl \leq L < \cl_E$ this equation
defines a closed curve centered about $(Q_+, M_1) = (0,1/4)$;
for increasing $L \rightarrow \cl_E$, the curves shrink to the
central point $(0,1/4)$. For $L =const > \cl_E$ the intersection
$(Q_0 =0) \cap (\ce = E) \cap (L = const)$ is empty. Hence,
an orbit with energy $E$ cannot cross $Q_0 = 0$
when $L>\cl_E$, which implies that $(Q_0 > 0)\cap (L > \cl_E)$
is a future invariant set with $L^\prime >0$
and $(Q_0 < 0)\cap (L > \cl_E)$
a past invariant set with $L^\prime < 0$.
From this the claim follows.
The statement can be strenghtened considerably:
there exists $\tilde{\cl}_E < \cl_E$ such that the
statement remains valid when $\cl_E$ is replaced by $\tilde{\cl}_E$.
The quantity $\tilde{\cl}_E$ is defined as the minimum of all $L=const$
such that $F < 0$ (and hence $Q_0^\prime < 0$)
on $(Q_0 =0) \cap (\ce = E) \cap (L =const)$
(for $F$ see Eq.~(\ref{decoup}));
in Fig.~\ref{semifig}, $L$ is chosen according to
$\cl_E > L =const > \tilde{\cl}_E$:
hence $(Q_0 =0) \cap (\ce = E) \cap (L =const)$ is not empty,
but it is contained in the region $F<0$.
By definition, for all orbits with energy $E$,
for all $L > \tilde{\cl}_E$, the hyperplane $Q_0 =0$
acts as a semipermeable membrane: orbits can only pass
through $Q_0 = 0$ from $Q_0 < 0$ to $Q_0 > 0$.
It follows that $(Q_0 > 0)\cap (L > \tilde{\cl}_E)$
is a future invariant set with $L^\prime >0$
and $(Q_0 < 0)\cap (L > \tilde{\cl}_E)$
a past invariant set with $L^\prime < 0$, and hence the claim is shown.
A very good approximation for the value of $\tilde{\cl}_E$
is given by the solution of the equation
$V(\tilde{\cl}_E)= E\, (4/3) (1-w(\tilde{\cl}_E))^{-1}$.

\section{H-normalized variables}
\label{H-norm}

We here give a comparison between the variables used in the paper
and the important H-normalized quantities used in
e.g.,~\cite{book:waiell97} (note that in contrast to our variables
$H$-normalized variables break down when $H=0$):
\begin{equation}\label{var}
\Sigma_+= \frac{Q_+}{Q_0}\, ,\quad N_1^2=\frac{12\,M_1}{Q_0^2}\,
,\quad N_2^2= \frac{3\,(1-Q_0)^2}{4\,M_1\,Q_0^2}\, ,
\end{equation}
\begin{equation}\label{quant}
\Omega_H  = \frac{\Omega_D}{Q_0^2}= \frac{1-Q_+^2-M_1}{Q_0^2}\,
,\quad q
=2\Sigma_+^2+\textfrac{1}{2}(1+3w)\Omega_H=\frac{F+Q_0Q_+}{Q_0^2}\,.
\end{equation}

\end{appendix}


\begin{thebibliography}{99}


\bibitem{book:waiell97}
J.~Wainwright and G.~F.~R.~Ellis.
\newblock {\em Dynamical systems in cosmology}.
\newblock {C}ambridge {U}niversity {P}ress, Cambridge, 1997.

\bibitem{uggmuhl990}
C.~Uggla and H. von Zur-M\"{u}hlen.
\newblock Class.\ Quantum Grav. \textbf{7}~:~1365--1385 (1990).

\bibitem{golell1999}
G.~F.~R.~Ellis and M.~Goliath.
\newblock Phys.\ Rev.\ D \textbf{60}~:~023502 (1999).

\bibitem{colgol2000}
A.~Coley and M.~Goliath.
\newblock Phys.\ Rev.\ D \textbf{62}~:~043526 (2000).

\bibitem{Lin/Wald:1990}
X.-f.~Lin and R.~M.~Wald.
\newblock Phys.\ Rev.\ D \textbf{41}~:~2444--2448 (1990).

\bibitem{Rendall:1995}
A.~Rendall.
\newblock Math.\ Proc.\ Camb.\ Phil.\ Soc. \textbf{118}~:~511--526 (1995).



\bibitem{Oliveira/etal:1997}
H.~P.~de Oliveira, I.~Dami\~ao Soares and T.~J.~Stuchi.
\newblock Phys.\ Rev.\ D \textbf{56}~:~730--740 (1997).

\bibitem{Barguine/etal:2001}
R.~Barguine, H.~P.~de Oliveira, I.~Dami\~ao Soares and E.~V.~Tonini.
\newblock Phys.\ Rev.\ D \textbf{63}~:~063502 (2001).


\bibitem{olietal2002}
H.~P.~de Oliveira, A.~M.~Ozorio de Almeida, I.~Dami\~ao Soares and E.~V.~Tonini.
\newblock Phys.\ Rev.\ D \textbf{65}~:~083511 (2002).


\bibitem{uggetal2004}C.~Uggla,, H.~van Elst, J.~Wainwright and G.~F.~R.~Ellis.
\newblock Phys.\ Rev.\ D \textbf{68}~:~103502 (2003).

\bibitem{ashetal1993}
A.~Ashtekar, R.~S.~Tate and C.~Uggla.
\newblock Int.\ J.\ Mod.\ Phys.\ D \textbf{2}~:~15--50 (1993).

\bibitem{heiugg2003}
J.~M.~Heinzle and C.~Uggla.
\newblock Ann.\ Phys. \textbf{308-1}~:~18--61 (2003).

\bibitem{heietal2003}
J.~M.~Heinzle, N.~R\"ohr and C.~Uggla.
\newblock Class.\ Quantum Grav. \textbf{20}~:~4567--4586 (2003).

\bibitem{book:gravitation}
C.~W.~Misner, K.~S.~Thorne and J.~A.~Wheeler.
\newblock {\em Gravitation}.
\newblock {W}.~{H}.~{F}reeman and {C}ompany, San Francisco, 1973.

\bibitem{cra1991}
J.~D.~Crawford.
\newblock Rev.\ Mod.\ Phys. \textbf{63-4}~:~991--1038 (1991).

\bibitem{stryor1967}
A.~Strauss and J.~A.~Yorke.
\newblock Math.\ Syst.\ Theory \textbf{1}~:~175--182 (1967).

\bibitem{horetal2003} J.~T.~Horwood, M.~J.~Hancock, D.~The and J.~Wainwright.
\newblock Class.\ Quantum Grav. \textbf{20}~:~1757--1778
(2003).

\bibitem{waietal1999}
J.~Wainwright, M.~J.~Hancock and C.~Uggla.
\newblock Class.\ Quantum Grav. \textbf{16}~:~2577--2598
(1999).



\end{thebibliography}
\end{document}